\documentclass[11pt]{article}
\usepackage[lmargin=1in,bmargin=.8in,tmargin=1in,rmargin=1in]{geometry}
\usepackage{amsmath,amssymb,amsfonts, amsthm}
\usepackage{graphicx}
\usepackage{float}
\usepackage{hyperref}
\usepackage{bm}
\usepackage{authblk}
\usepackage[utf8]{inputenc}
\usepackage{csquotes}
\usepackage{caption}
\usepackage{subcaption}
\usepackage{xcolor}
\usepackage[capitalise]{cleveref}
\usepackage{url}
\usepackage{paralist}
\usepackage[switch]{lineno}
\usepackage{booktabs}

\usepackage{xr-hyper}
\usepackage[backend=biber, style=numeric-comp, sorting=none, giveninits=true, uniquename=false, backref=false, defernumbers=true,language=auto]{biblatex}
\begin{document}

\title{Predicting success of cooperators across arbitrary heterogeneous environmental landscapes}
\author[1]{Amir Kargaran}
\author[2]{Kamran Kaveh}
\author[3]{Krishnendu Chatterjee}

\affil[1]{School of Biological Sciences, Institute for Research in Fundamental Sciences (IPM), Tehran, Iran}
\affil[2]{University of Minnesota, Minneapolis, MN, USA}
\affil[3]{Institute of Science and Technology Austria, Klosterneuburg, Austria }

\maketitle

\begin{abstract}
Cooperation is central to the organization of complex biological and social systems. Most theoretical models assume homogeneous environments; in reality, populations inhabit spatially varying landscapes in which the payoffs of cooperation differ across space. Here, we introduce a general framework for the evolution of cooperation in complex, heterogeneous environments where the benefit of cooperation depends on local environmental quality. Cooperators in environmentally rich sites confer greater benefits than those on poor sites. We show that whether heterogeneity promotes or suppresses cooperation is determined primarily by the spatial organization of environmental states. Across arbitrary environmental landscapes, a single quantity, the spatial correlation index (SCI), predicts the fixation probability of cooperators. Under weak selection, segregated environments enhance cooperation, whereas highly intermixed, checkerboard-like landscapes suppress it. Beyond fixation probabilities, environmental organization also controls evolutionary timescales: segregated landscapes generate long-lived metastable coexistence, whereas intermixed landscapes lead to faster but less successful fixation of cooperators. Together, these results provide a unifying description of how spatial environmental heterogeneity shapes the evolution of cooperation and suggest measurable predictors of cooperative success in biological and social settings.
\end{abstract}


\section{Introduction}
Cooperation is a fundamental aspect of biological systems, manifesting across levels of complexity, ranging from genes and single-celled organisms to advanced human societies. Despite evolutionary pressures that typically favor selfish strategies that maximize individual fitness, cooperative and altruistic interactions are widespread. Their persistence suggests that cooperation provides advantages that enhance survival and group functionality \cite{nowak2006evolutionary,hamilton1964genetical,axelrod1981evolution,trivers1971evolution,
fehr2004social,Boyd2009Culture,Rand2013,nowak2011supercooperators}.

Evolutionary game theory provides a natural framework for studying cooperation, often through variants of the Prisoner's Dilemma (PD) game \cite{Smith1988,axelrod1981evolution,weibull1997evolutionary,hofbauer2003evolutionary,hofbauer1998evolutionary,broom2014game,doebeli2005models}. While in well-mixed populations, defection dominates, structured populations can promote cooperation. In particular, cooperators can prevail when the benefit-to-cost ratio exceeds a threshold determined by the topology of the population structure -- the $b/c$-rule \cite{Ohtsuki2006Nature,Nowak2006FiveRules,allen2017evolutionary,allenevolutionary2019,mcAvoyfixation2021,Fotouhi2019Royal,Fotouhi2018NatureHuman,traulsen2009stochastic,hauert2004spatial,allen2014games,su2022evolution,fu2009evolutionary}. These results demonstrate how the population structure shapes cooperative social behavior.

\begin{figure}
\centering
\includegraphics[width=.8\linewidth]{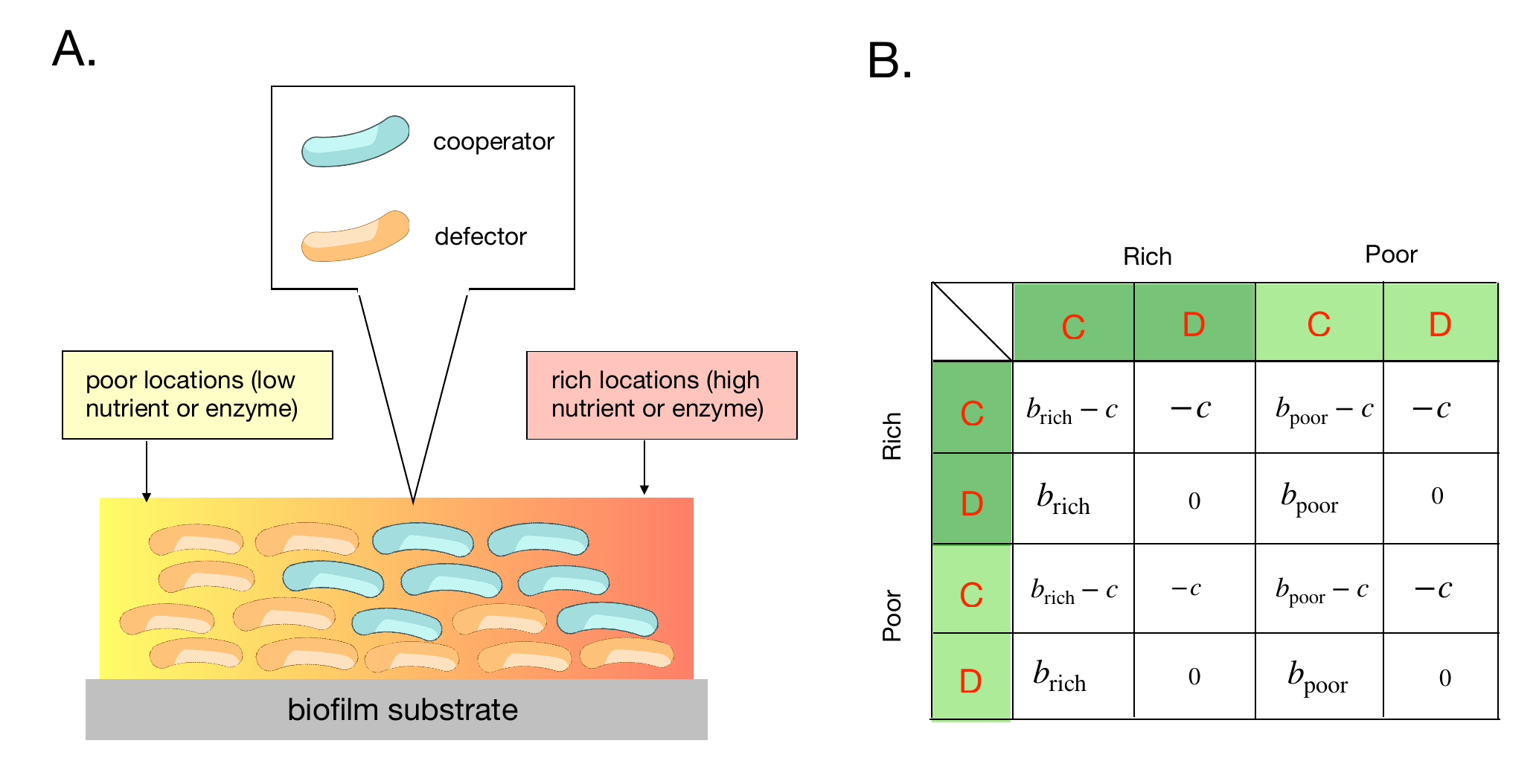}
\caption{\textbf{Schematics of cooperation in microbial biofilms and the extended payoff matrix.} Schematic illustration of cooperation in microbial systems, such as public-goods production in Pseudomonas (siderophore secretion) or predation in \textit{Myxococcus Xanthus}, among many other examples \cite{buckling2007siderophore,thiery2020predation,damore2012understanding,stilwell2020resource,contreras2024myxococcus}. {\bf B:} Payoff structure for cooperator-defector interactions in poor and rich environmental states, where cooperators incur a cost $c$ and provide benefits that depend on local environmental quality.}
\label{fig:main-fig1}
\end{figure}

However, real biological environments are rarely homogeneous. Microbial communities frequently inhabit spatially heterogeneous landscapes in which resource availability, metabolite diffusion, and the local microenvironment vary across space \cite{damore2012understanding,celiker2013cellular}. In \textit{Pseudomonas} biofilms, for example, siderophore-producing cooperators provide benefits that depend sensitively on local iron abundance \cite{nadell2016spatial,buckling2007siderophore,stilwell2020resource}. Similarly, in the social bacterium \textit{Myxococcus Xanthus}, the reward from cooperative predation depends on density and patchiness \cite{thiery2020predation,contreras2024myxococcus}. Cooperative behavior appears in numerous other biological systems, including cross-feeding microbes and tumor cell populations, where environmental factors such as nutrient gradients and metabolic patchiness modulate cooperative interactions \cite{le2024optogenetic,lee2023clonal,axelrod2006evolution,archetti2019cooperation}. 
Similarly, in human societies, cooperation is shaped by environmental effects: field studies show that resource variability, neighborhood inequality, and spatially uneven risk strongly modulate cooperative behavior in public-goods and collective-action settings \cite{ostrom1990governing,nishi2015inequality,henrich2006cooperation}. Across these diverse systems, environmental heterogeneity leads to spatial variation in the benefits that cooperators can confer. (Figure \ref{fig:main-fig1}).

Prior theoretical work on the effect of environmental heterogeneity in structured populations has focused primarily on constant-selection models \cite{kaveh2019environmental,kaveh2020moran,mahdipour2017genotype,farhang2017effect,farhang2019environmental,maciejewski2014environmental,Gavrilatzlets2002fixation,masuda2010heterogeneous,whitlock2005probability,manem2015modeling}, whereas game-theoretic interactions that incorporate environmental or ecological variations have received less attention. Existing studies typically explore restricted subsets of ecological or spatial configurations. In public-goods game models, random environmental heterogeneity is shown to stabilize coexistence or modulate cooperation \cite{Kun2013,stilwell2020resource}, whereas in some group-interaction settings, heterogeneity has been argued not necessarily to promote cooperation \cite{perc2011does}. Other models incorporate non-spatial forms of payoff variation, such as environmental feedback or temporal or strategy heterogeneity \cite{szolnoki2018environmental,amaral2020heterogeneity,SantosPNAS}. However, these approaches generally address particular types of heterogeneity, often uncorrelated or randomly assigned, and do not generalize to the broader space of possible environmental landscapes. As a result, we lack a framework that links a given spatial organization of payoffs to evolutionary outcomes. In particular, no existing work provides a unifying predictor for when any given heterogeneous ecological landscape will promote or suppress cooperation.

Here, we introduce a general and predictive model for the evolution of cooperation in spatially quenched heterogeneous environments, helping to unify our understanding of how environmental structure shapes evolutionary dynamics. We study a donation game on 1D and 2D lattice graphs, in which cooperative benefits vary across space. Cooperators in ``rich'' sites provide greater benefit than those in ``poor'' sites. To capture the essential features of such landscapes, we define two quantities: (i) the 
{\it heterogeneity magnitude} $\sigma$, which sets the payoff contrast between sites, and (ii) 
the {\it spatial correlation index} (SCI), a single scalar descriptor that quantifies 
spatial clustering or mixing of environmental states (Figure \ref{fig:main-fig2}).

We find that the spatial topology of environmental states, summarized by the
SCI, determines whether heterogeneity promotes or suppresses cooperation. In particular, segregated landscapes amplify cooperation as $\sigma$ increases, whereas highly intermixed (checkerboard-like) configurations suppress cooperation. Meanwhile, the classical selection condition, the $b/c$-rule, remains robust.

Across fully arbitrary spatial configurations, the SCI accurately predicts the fixation probability, providing the first unifying measure of how heterogeneous payoff landscapes shape cooperative success: fixation probabilities vary systematically, but modestly, across configurations, consistent with the weak-selection regime. Importantly, the spatial organization of the environment also regulates evolutionary timescales: segregated environments give rise to long-lived metastable coexistence, whereas highly intermixed environments lead to faster fixation dynamics.

    
\section{Model}\label{section two}
We consider a simplified two-player Prisoner's Dilemma game with two strategies: cooperators (C) and defectors (D). In a uniform environment, a cooperator interacting with another player provides a benefit $b$ at a personal cost $c$, while defectors incur no costs and provide no benefits. Fitness is linearly derived from payoff, scaled by a selection-intensity parameter $w$ $(0 < w < 1)$ (see Supplementary Note 1). We consider lattice graph population structures: a one-dimensional cycle with degree $k = 2$ and a two-dimensional square lattice with degree $k = 4$. Evolutionary dynamics follows a spatial Moran death–birth process \cite{Moran_1958, lieberman2005evolutionary}. The fixation probability $\rho_{C}$ of a single randomly placed cooperator is our primary measure of cooperative success.\\
    
{\bf Payoffs and the environmental landscape.} Environmental heterogeneity is introduced by assigning a rich or poor condition to each node, affecting the benefit of cooperation while keeping the cost constant. Cooperators in poor locations provide benefit, $b_{\text{poor}}$ and in rich locations, $b_{\text{rich}}$. We define
\begin{align}
    b_{\text{poor}} = b_{\text{ave}} - \sigma, \nonumber\\
    b_{\text{rich}} = b_{\text{ave}} + \sigma,  \nonumber
\end{align}
where $b_{\text{ave}} = (b_{\text{rich}} + b_{\text{poor}})/2$ and $\sigma = (b_{\text{rich}} - b_{\text{poor}})/2$ represent the mean benefit and half-range. We refer to $\sigma$ as the `heterogeneity magnitude'. To ensure non-negative payoffs in poor environments, we require $b_{\text{poor}} \ge 0$, which implies $\sigma / b_{\text{ave}} \le 1$. We define the normalized heterogeneity parameter $\sigma / \sigma_{\text{max}}$, where $\sigma_{\text{max}} = b_{\text{ave}}$. Thus, $\sigma / \sigma_{\text{max}} = 1$ represents maximal heterogeneity and $\sigma / \sigma_{\text{max}} = 0$ corresponds to a homogeneous environment. We assume equal numbers of rich and poor sites so that the mean benefit across the lattice is $b_{\text{ave}}$. Each node’s condition is encoded in a configuration vector $\bm{\mathcal{V}} = (1,1,1,-1,-1,1,\ldots)$, where $+1$ indicates a rich node and $-1$ a poor node. The benefit vector is:
\begin{align}
    \bm{b} = b_{\text{ave}} + \sigma\, \bm{\mathcal{V}}.
\end{align}
    
\begin{figure}
\centering
\includegraphics[width=.8\linewidth]{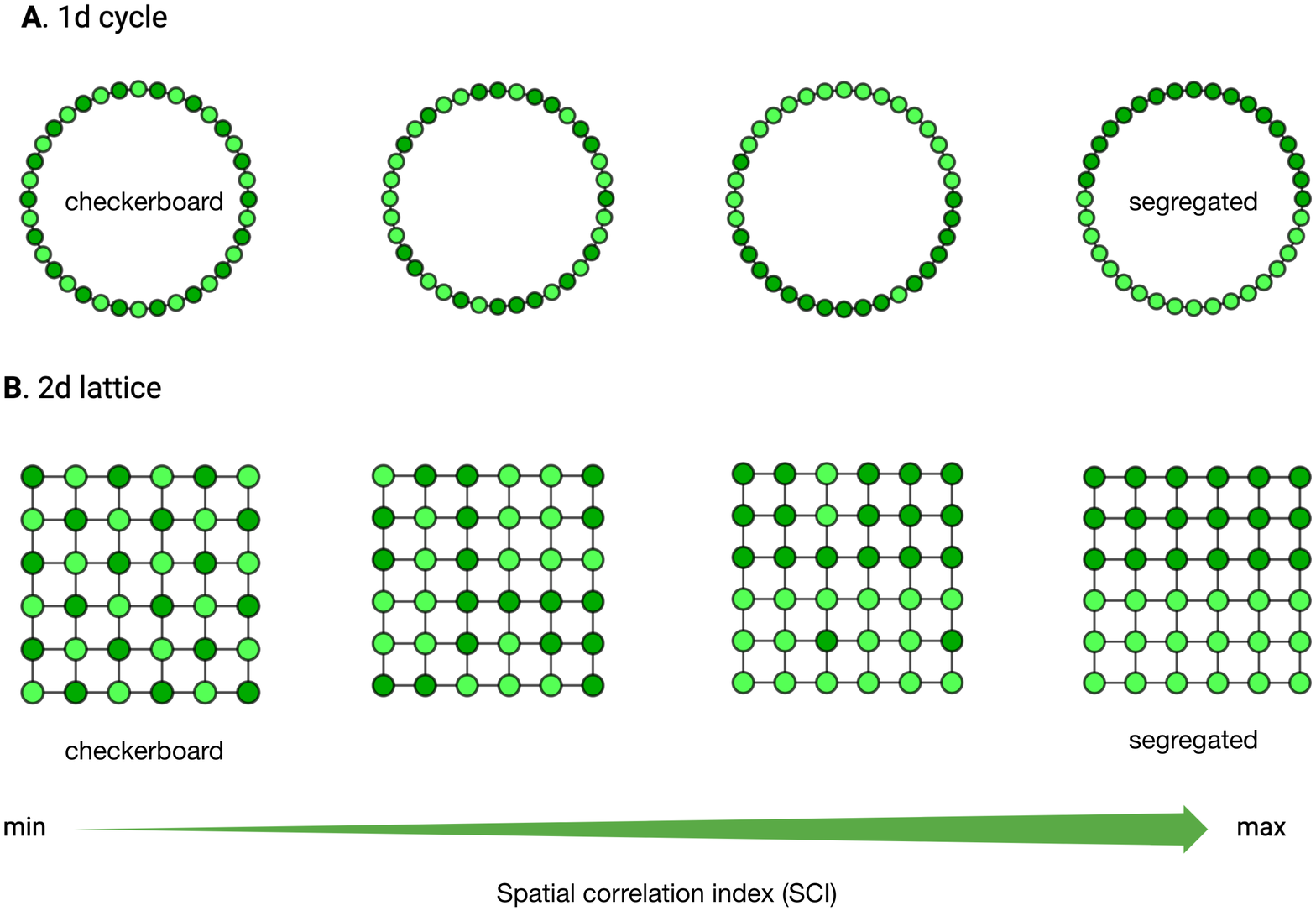}
\caption{\textbf{Environmental configurations with varying spatial correlation (SCI)} {\bf A:} 1D cycle graph with arrangements ranging from \enquote{checkerboard} to \enquote{segregated}. {\bf B:} analogous 2D lattice configurations. Periodic boundaries are assumed. These examples span the full range of spatial correlation index (SCI), with two intermediate random configurations illustrated. (Dark green indicates `rich' and light green indicates `poor' locations)}
\label{fig:main-fig2}
\end{figure}

We are interested in how the properties of $\bm{\mathcal{V}}$, particularly the degree of mixing or clustering of environmental states, influence cooperation. Two illustrative configurations are: (i) \emph{checkerboard (2-chromatic)}: each poor node is surrounded by rich neighbors, and vice versa; (ii) \emph{segregated}: rich and poor nodes form large separate spatial clusters with minimal mixing between the two (Figure \ref{fig:main-fig2}).\\

{\bf Spatial correlation index (SCI).} To quantify the spatial organization of environmental states, we introduce the spatial correlation index (SCI):
\begin{equation}
    \text{SCI} = \sum_{i=1}^{N}\sum_{j=1}^{i} \frac{v_i v_j}{\delta(i,j)^\alpha},
\label{SCI_index}
\end{equation}
where $\delta(i,j)$ is the distance between nodes $i$, $j$ and $v_i,v_j$ are the elements of $\bm{\mathcal{V}}$. The exponent $\alpha = 1/2$ and is fixed throughout. The SCI is a distance-weighted two-point correlation function: it increases when similar environmental states cluster and decreases when rich and poor sites alternate. The value of $\alpha = 1/2$ was chosen for clarity, but the results are robust for a broad range of exponents and kernel shapes. Any reasonable kernel yields the same qualitative ordering of landscapes and preserves the relationship between SCI and fixation probability (see next section). Thus, SCI reflects an underlying correlation structure rather than a specific mathematical choice.\\
    
{\bf Measures of success for cooperators:} Our goal is to predict the fixation probability of a cooperator,
\[\rho_C = \rho(b_{\rm ave}, c, \sigma, \bm{\mathcal{V}}),\]
for different levels of heterogeneity $\sigma$ and spatial arrangement $\bm{\mathcal{V}}$. Cooperator strategy is advantageous if $\rho_D < 1/N < \rho_C$ \cite{Ohtsuki2006Royal}. ($\rho_D$ is the respective fixation probability of the defectors.)
    
As a secondary measure, we consider the {\it level of cooperation}, $f_c$, or the long-term, but finite, fraction of cooperators. This provides insight into how environmental configurations affect the dynamics and timescales of cooperative fixation.\\

{\bf Simulation protocol.} Fixation probabilities were calculated by stochastic simulation of the
death–birth Moran process on 1D and 2D lattices with periodic boundary conditions. For each configuration, we ran between
$10^7-10^8$ stochastic realizations until absorption (full cooperation or full
defection), and the fixation probability was computed as the fraction of runs ending in
cooperators' fixation. The error bars obtained were negligible. For the long-term cooperation level, we initialized populations with $10\%$ cooperators and simulated $100$ trajectories up to $t = 10^9$, averaging only over runs in
which cooperation persisted. Randomly mixed configurations were generated by applying a
prescribed number of pairwise swaps to the segregated state, producing Lightly-Mixed,
Moderately-Mixed, and Well-Mixed regimes.

        
\section{Results}\label{section three}
\subsection{Low environmental mixing facilitates cooperation (segregated configuration)}
Figure \ref{fig:main-fig3}(d, e) shows heat maps of the normalized fixation probability for the segregated configuration, plotted against the average benefit-to-cost ratio ($b_{\mathrm{ave}}/c$) and the normalized heterogeneity ($\sigma/\sigma_{\text{max}}$) in both the cycle graph (1D) and the square lattice (2D). In both population structures, increasing either parameter raises the fixation probability of cooperators, with a stronger effect in the square lattice. The normalized fixation probability is defined as $\rho/\rho_0$, where $\rho_0$ is the fixation probability in a homogeneous environment ($\sigma = 0$).
    
The panels (a–c, f) provide horizontal and vertical cross-sections of these heat maps. In all cases, the homogeneous baseline ($\rho/\rho_0 = 1$) is indicated by the black dashed line. For moderate and maximal heterogeneity ($\sigma/\sigma_{\text{max}} = 0.5, 1$), the fixation probability consistently exceeds this baseline: at full heterogeneity, it is approximately 8\% higher in the cycle graph and 4\% higher in the square lattice. Notice that the magnitude of the change in fixation probability, while modest, is comparable to that reported under weak selection for heterogeneous graphs relative to regular structures \cite{mcAvoyfixation2021,ohtsuki2006simple}. Supplementary Note 4 presents an approximate analytical description of this trend and an approximate \textit{b/c}-rule for the cycle graph. 

\subsection{High environmental mixing suppresses cooperation (checkerboard configuration)}
Figure \ref{fig:main-fig4}(d, e) shows heat maps of the normalized fixation probability for the checkerboard (2-chromatic) configuration, plotted against normalized heterogeneity ($\sigma/\sigma_{\text{max}}$) and the average benefit-to-cost ratio ($b_{\mathrm{ave}}/c$) in both the cycle graph and the square lattice. In contrast to the segregated case, increasing heterogeneity now decreases the fixation probability, with similar magnitudes of decline in 1D and 2D.
    
The panels (a–c, f) show horizontal and vertical cross-sections of the heat maps. The homogeneous baseline ($\rho/\rho_0 = 1$) is shown as a black dashed line. For moderate and maximal heterogeneity ($\sigma/\sigma_{\text{max}} = 0.5, 1$), the fixation probabilities fall below this baseline: at full heterogeneity, they are reduced by approximately 5\% in both network types. Panels (c) and (f) reveal the same declining trend at fixed $b_{\mathrm{ave}}/c$. 
    
Overall, the checkerboard configuration suppresses cooperation, making cooperators more likely to go extinct than in a homogeneous environment. Supplementary Notes 2 and 3 provide, respectively, exact fixation-probability derivation and an approximate \textit{b/c}-rule for the checkerboard configuration in the cycle graph.

\subsection{Spatial correlation of payoffs (benefits) is the primary determinant of cooperation}
Figure \ref{fig:main-fig5} shows the normalized fixation probability for a large set of random and periodic configurations, as well as limiting configurations, segregated and checkerboard, each plotted as a function of the spatial correlation index (SCI) on the cycle graph (panel a) and the square lattice (panel b). Random configurations are generated by applying a random number of swaps to the segregated benefit vector. For example, Well-Mixed configurations use a randomly chosen number of swaps between $10^2$ and $10^3$, while Moderately Mixed and Lightly Mixed configurations use random numbers in the ranges $10$–$10^2$ and $1$–$10$, respectively.
    
Periodic configurations (labeled “Periodic I’’ and “Periodic II’’) are constructed so that rich and poor sites alternate with fixed intervals. On the cycle graph, in the second periodic configuration, the sites $1$–$25$ are assigned as poor and the sites $26$–$50$ as rich, and this pattern repeats. In square lattices, periodicity is achieved by alternately assigning benefit types to rows and columns.
    
Across all configurations, segregated networks produce the highest normalized fixation probability, while checkerboard configurations produce the lowest. A linear fit through the intermediate configurations shows that the slope in the square lattice is roughly half that of the cycle graph. Although this fitted line predicts the fixation probability of the segregated configuration, it does not accurately capture the fixation probabilities of the checkerboard or periodic configurations.
    
Schematic examples of each configuration are also shown in Figure \ref{fig:main-fig5} and illustrate how the spatial correlation index (SCI) reflects the arrangement of heterogeneous sites. Supplementary Notes 5 further analyze how the fixation probability varies with the heterogeneity parameter in random, periodic, and pattern-like configurations. 

\subsection{\texorpdfstring{The $b/c$-rule is largely unaffected by environmental configuration}{The b/c-rule is largely unaffected by environmental configuration}}

\begin{figure}
\centering
\includegraphics[width=1\linewidth]{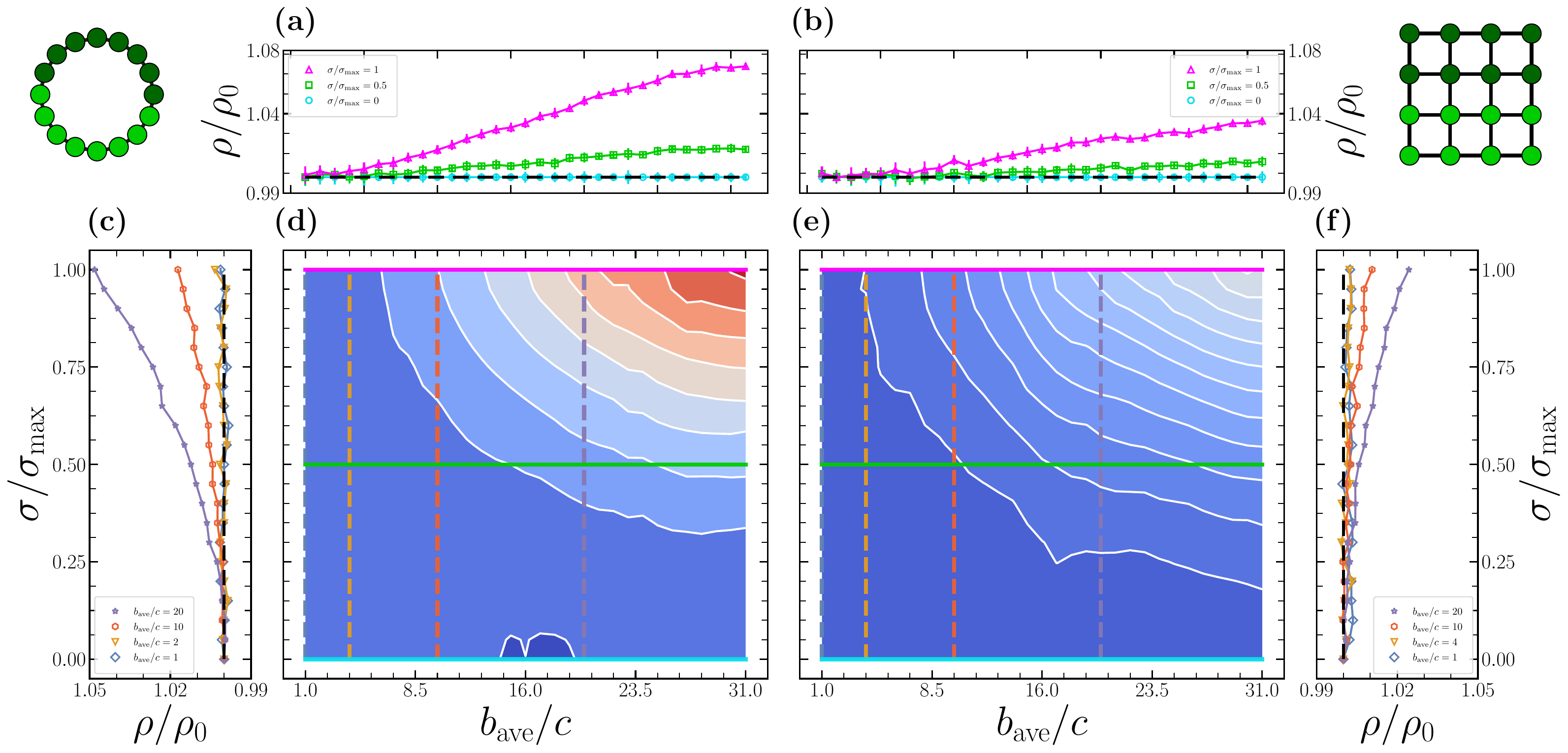}
\caption{\textbf{Normalized fixation probability (\(\rho/\rho_0\)) for segregated configurations.} heat maps (panels d, e) show \(\rho/\rho_0\) as a function of normalized heterogeneity \(\sigma/\sigma_{\text{max}}\) and the average benefit-to-cost ratio \(b_{\text{ave}}/c\). The cross-sections for fixed \(\sigma/\sigma_{\text{max}}\) (panels a, b) and fixed \(b_{\text{ave}}/c\) (panels c, f) highlight trends. \(\rho/\rho_0 > 1\) (above black dashed line) indicates that cooperation is modestly amplified by heterogeneity. Simulations use \(N = 100\) (1D) and \(10 \times 10\) (2D) lattices, with \(c = 0.125\). The insets  show schematic examples (cropped for clarity).}
\label{fig:main-fig3}
\end{figure}

\begin{figure}
\centering
\includegraphics[width=1\linewidth]{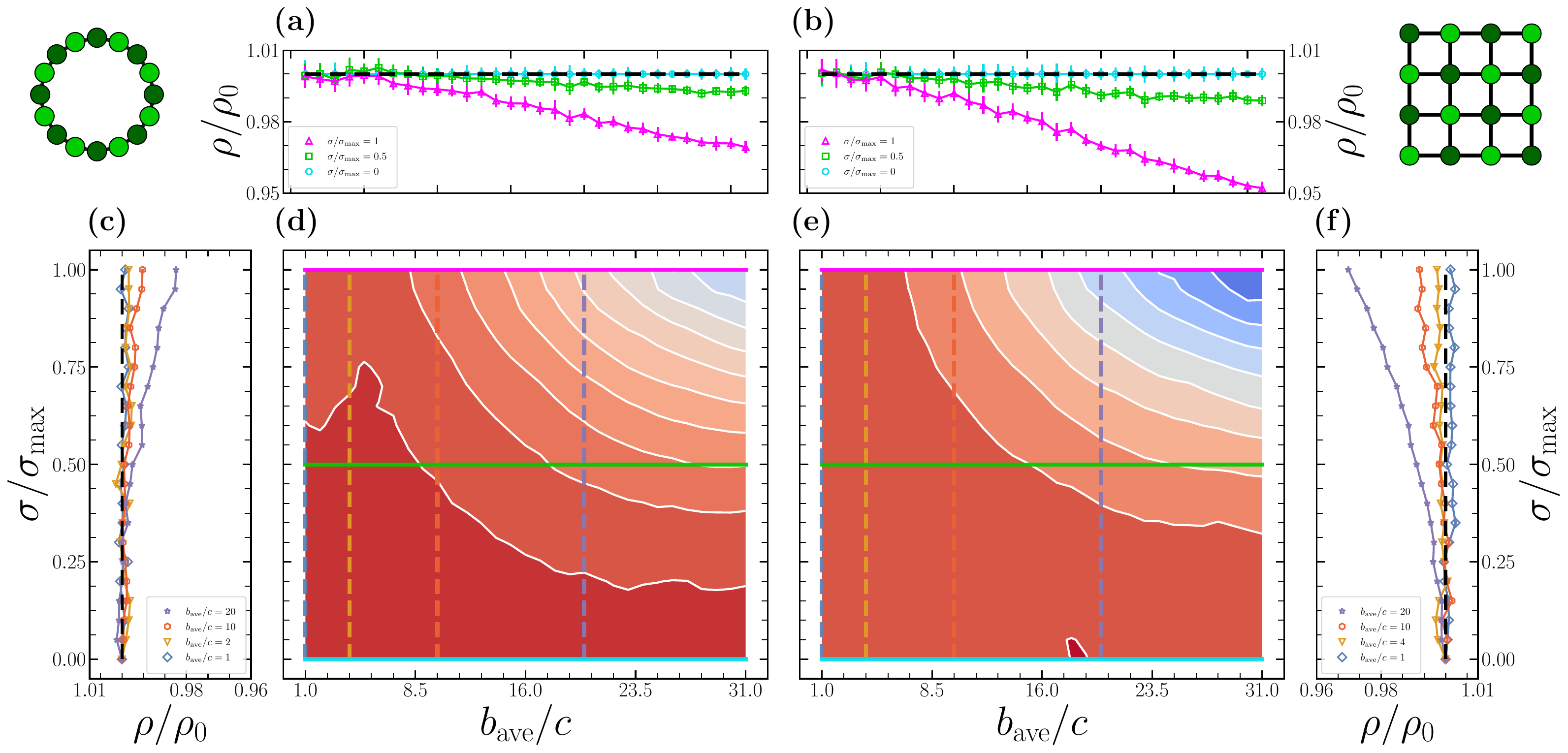}
\caption{\textbf{Normalized fixation probability (\(\rho/\rho_0\)) for checkerboard configurations.} heat maps (d, e) show \(\rho/\rho_0\) as a function of normalized heterogeneity \(\sigma/\sigma_{\text{max}}\) and the benefit-to-cost ratio \(b_{\text{ave}}/c\). Panels (a, b) show horizontal cross-sections at fixed \(\sigma/\sigma_{\text{max}}\); panels (c, f) show vertical cross-sections at fixed \(b_{\text{ave}}/c\). Fixation probability decreases with increasing heterogeneity. The black dashed line (\(\rho/\rho_0 = 1\)) indicates the neutral threshold. Simulations use a 1D lattice with \(N=100\) and a \(10 \times 10\) 2D lattice with \(c = 0.125\). Insets depict simplified schematics (16 nodes are shown for better visualization).}
\label{fig:main-fig4}
\end{figure}

\begin{figure}
\centering
\includegraphics[width=.8\linewidth]{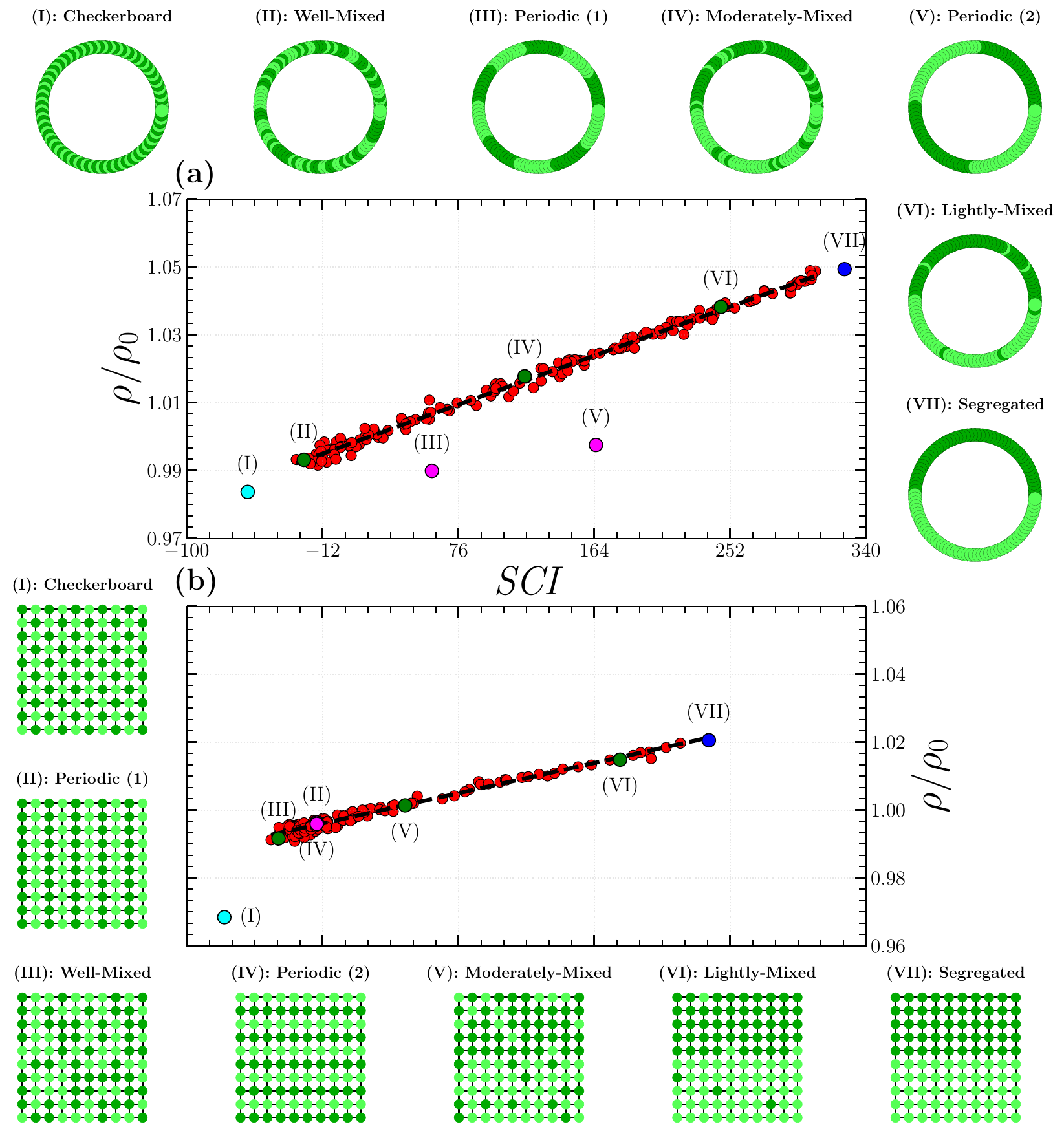}
\caption{\textbf{Normalized fixation probability versus spatial correlation index (SCI)} for cycle graphs (a) and square lattices (b) with 100 nodes ($N=100$; $10\times10$), with high heterogeneity ($b_{\mathrm{ave}}/c = 20$, $\sigma/\sigma_{\text{max}} = 1$). The red points denote 150 random configurations, and the three green points correspond to specific random classes (\enquote{Well-Mixed}, \enquote{Moderately Mixed}, \enquote{Lightly Mixed}). Magenta points mark two periodic configurations (\enquote{Periodic I}, \enquote{Periodic II}); in 1D these use periods 10 and 25, and in 2D they are generated by alternating rows and columns. Segregated and checkerboard configurations are shown in blue and cyan, respectively, representing the two extremes of SCI. Under weak selection, the normalized fixation probability varies nearly linearly with SCI, as indicated by the fitted black dashed lines. The slope in the square lattice is nearly half that of the cycle graph. The error bars for all points are below $4 \times 10^{-3}$. Simulations use $c = 0.125$.}
\label{fig:main-fig5}
\end{figure}

Heterogeneity changes the fixation probability, but has little effect on the condition for selection. That is, the selection condition for cooperators under weak selection is independent of the heterogeneity amplitude $\sigma$. As shown in Figures \ref{fig:main-fig3} and \ref{fig:main-fig4}, for $b/c = 1$ (in 1D) and $b/c = 4$ (in 2D), the normalized fixation probability 
$\rho/\rho_0$ remains close to 1 for all values of $\sigma$. 

For uniform environments, Ohtsuki et al.~\cite{Ohtsuki2006Royal} showed that the classical 
\textit{b/c}-rule arises from the first-order term in the weak-selection expansion of the 
fixation probability. Using the same type of expansion for both the segregated 
(Supplementary Note 4) and checkerboard (Supplementary Note 3) configurations, we find the 
same result: under weak selection ($w \ll 1$), the first-order term, and therefore the 
\textit{b/c}-rule, remains unchanged by environmental heterogeneity. Only higher-order terms 
depend on the spatial organization of environmental states.

\subsection{Environmental mixing regulates fixation times and metastable coexistence}

We now turn to a secondary measure of cooperative success: the level of cooperation ($f_c$), defined as the long-term fraction of cooperators. We track the temporal evolution of $f_c$ starting from an initial condition with $10\%$ cooperators placed uniformly at random, for both checkerboard and segregated configurations on one- and two-dimensional lattices. Each panel in Fig.~\ref{fig:main-fig6} shows averages over 100 runs in which cooperation persists, for different values of $b_{\mathrm{ave}}$ and $\sigma$.

Panels (a) and (c) show the long-term behavior of the checkerboard configuration. As in homogeneous environments, $f_c$ exhibits a sharp transition: once the benefit-to-cost ratio exceeds a critical threshold, the system rapidly converges to full cooperation. In contrast, the segregated configuration (panels (b) and (d)) displays qualitatively different behavior. Even when $b_{\mathrm{ave}}/c$ exceeds the corresponding threshold, the fraction of cooperators often fails to reach unity within the simulated time window, indicating a dramatic slowdown of the fixation dynamics induced by environmental heterogeneity.

This effect is particularly pronounced in one dimension. For sufficiently large heterogeneity ($\sigma/\sigma_{\max}$) and even for $b_{\mathrm{ave}}/c \sim 8$--16, the system becomes trapped in long-lived partially cooperative states with $f_c \approx 1/2$ (beige regions). This behavior is reminiscent of localization phenomena in disordered environments, where spatial heterogeneity inhibits global spread \cite{Sinai,sinai1982limit,hughes1996random,masuda2010heterogeneous}.

To further characterize these dynamics, we analyzed temporal fixation trajectories for additional mixing levels (Figs.~S9--S10). At short times ($t < 10^8$), all configurations exhibit similar growth. At longer times, however, the dynamics diverge sharply: checkerboard configurations approach full cooperation rapidly, whereas segregated configurations relax extremely slowly and often plateau, rendering fixation effectively unattainable on biologically relevant timescales.

Taken together, these results show that while environmental heterogeneity may only modestly alter fixation probabilities, it can profoundly reshape evolutionary timescales. Segregated landscapes promote cooperation in principle, yet simultaneously generate metastable coexistence that delays fixation by orders of magnitude.


\section{Discussion}

We introduced a general framework for understanding how environmental heterogeneity shapes the evolution of cooperation in structured populations. While population structure is known to influence cooperative success \cite{Ohtsuki2006Nature,Nowak2006FiveRules,allen2017evolutionary}, the role of heterogeneous ecological landscapes has remained unclear. By embedding a spatially varying benefit field into a Moran death-birth process \cite{Moran_1958,lieberman2005evolutionary}, we show that the spatial organization of environmental states, not merely the magnitude of heterogeneity, plays a decisive role in determining cooperative outcomes.

Our results reveal two distinct evolutionary regimes. Highly intermixed environments, exemplified by the checkerboard configuration, are associated with reduced cooperative success, whereas segregated environments are associated with higher cooperative success. These configurations define the extremes of the observed behavior, with intermediate landscapes smoothly interpolating between them. A single scalar measure, the spatial correlation index (SCI), captures this continuum: the fixation probability varies almost linearly with SCI across random, periodic, and patterned landscapes in one- and two-dimensional systems. Thus, SCI serves as a unifying predictor of cooperative success across arbitrary heterogeneous landscapes. Conceptually, SCI plays a role analogous to `modularity' in network science \cite{newman2006modularity,girvan2002community,newman2004finding}.

Despite its influence on fixation probabilities, environmental heterogeneity does not alter the classical weak-selection condition for cooperation. The $b/c$-rule \cite{Ohtsuki2006Nature,Nowak2006FiveRules,allen2014games,allen2017evolutionary} emerges from the first-order term in the weak-selection expansion, which is unaffected by the spatial arrangement of payoffs; only higher-order terms depend on environmental organization (Supplementary Notes 2 and 4). Thus, heterogeneity modulates the extent and timing of cooperative success without shifting the threshold at which selection begins to favor cooperation.

Environmental structure also strongly influences the evolutionary dynamics. Intermixed environments lead to rapid absorption with little delay, whereas segregated environments generate much slower dynamics characterized by long-lived metastable coexistence (Supplementary Note 6). These contrasting timescales demonstrate that environmental topology shapes not only the direction but also the tempo of evolution.

Although our framework focuses on one- and two-dimensional lattices for clarity and tractability, the SCI framework is not lattice-specific. SCI is defined purely through pairwise correlations between environmental states and a graph-distance metric, and therefore, generalizes naturally to arbitrary population structures, including irregular graphs, weighted networks, and empirical spatial embeddings.

Similarly, while we employed a bimodal environmental distribution (rich/poor) for conceptual simplicity, the SCI framework extends directly to continuous or multi-level environmental variation. In such settings, environmental states become real-valued, and SCI reduces to a weighted spatial autocorrelation measure analogous to two-point correlation functions used in statistical physics and landscape ecology. Because the fixation probability depends on spatial correlations rather than on the number of environmental classes, the predictive capacity of SCI is expected to extend to continuous landscapes.

\begin{figure}
\centering
\includegraphics[width=1\linewidth]{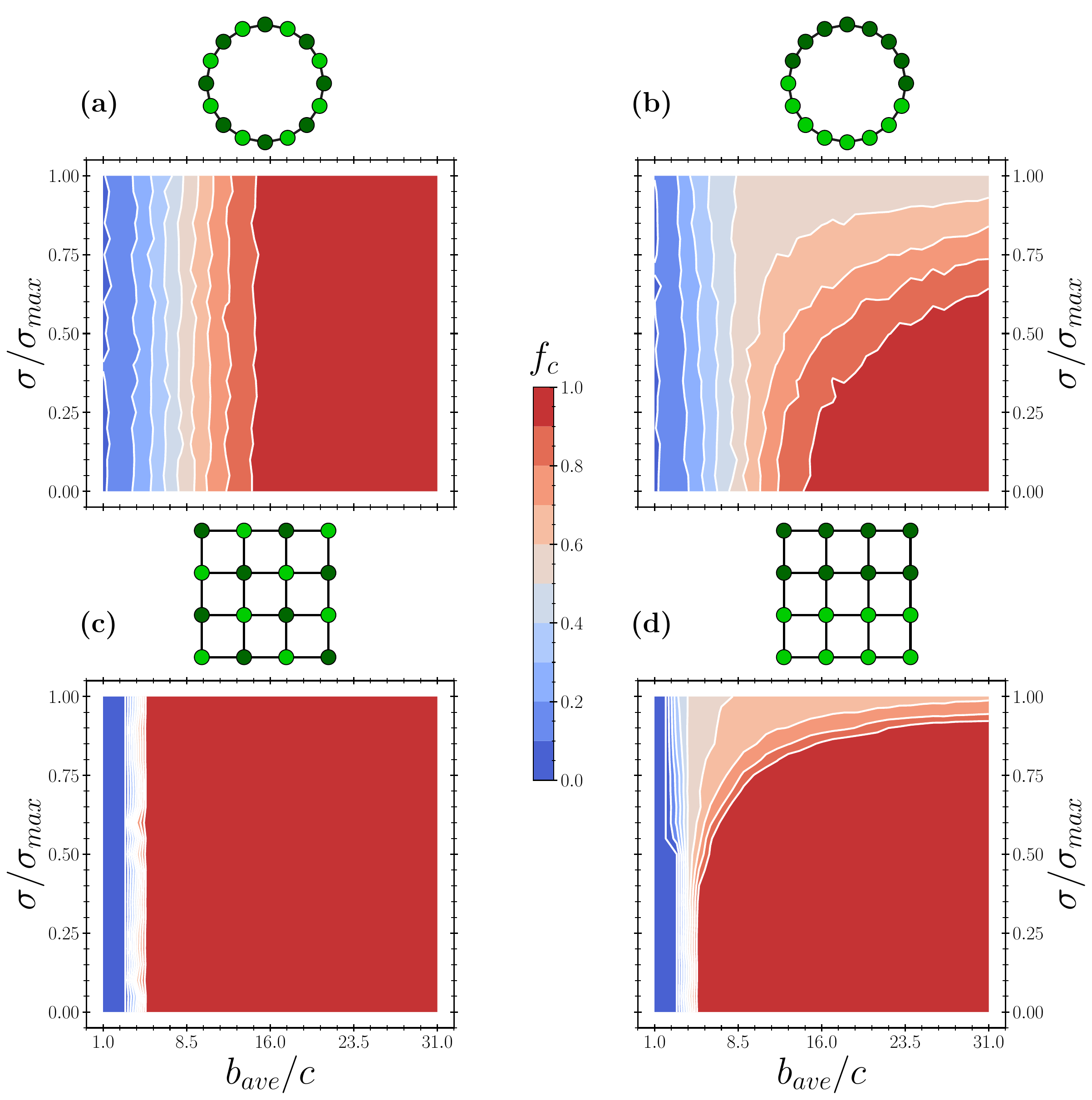}
\caption{\textbf{Spatial organization of environmental states regulates fixation dynamics.} Panels (a,c) show the long-term level of cooperation ($f_c$) for the checkerboard configuration on a cycle graph ($N=10^4$) and a square lattice ($150 \times 150$), respectively; panels (b,d) show the segregated configuration. Populations are initialized with $10\%$ cooperators. The results are shown for $t= 10^9$ as heat maps as both $b_{\mathrm{ave}}$ and $\sigma$ are varied. Checkerboard configurations exhibit a sharp transition to fixation, whereas segregated configurations display long-lived metastable coexistence.}
\label{fig:main-fig6}
\end{figure}

As expected under weak selection, changes in fixation probability are modest, and of a magnitude comparable to those reported for heterogeneous population structures under weak selection \cite{mcAvoyfixation2021,ohtsuki2006simple}, however, even in this regime, the effects on evolutionary timescales can be substantial (Figure~\ref{fig:main-fig6}). Segregated environments give rise to long-lived metastable states, whereas intermixed environments do not affect fixation time. These differences in timescales have important biological implications, particularly for large populations or systems subject to repeated invasion events, where evolutionary outcomes depend sensitively on the persistence of transient coexistence.

Spatial heterogeneity is pervasive in biological systems. Microbial communities experience patchy nutrient distributions \cite{damore2012understanding,celiker2013cellular}, and cooperative behaviors such as siderophore production \cite{nadell2016spatial,buckling2007siderophore,stilwell2020resource} and \textit{Myxococcus} predation \cite{thiery2020predation,contreras2024myxococcus} depend critically on the local microenvironment. Analogous principles arise in tumor ecology \cite{lee2023clonal,archetti2019cooperation} and in human social systems shaped by spatial inequality \cite{ostrom1990governing,nishi2015inequality}. For such systems, SCI-like correlation measures, inferred from imaging, spatial transcriptomic, or socioeconomic data, may offer practical predictors of cooperative outcomes.

In summary, we identify SCI as a simple and predictive descriptor of cooperative success across heterogeneous spatial landscapes. By linking the environmental topology with the fixation probabilities, the selection thresholds, and the evolutionary timescales, our framework provides a unifying perspective on cooperation in spatially structured ecological and social systems.

    
\printbibliography[]
    
\newpage
    
\clearpage
\section*{Acknowledgements}
The authors thank M. Sadeghi for his valuable discussions and insights that significantly contributed to this work. We also thank Turin Cloud Services for providing the computational resources necessary for conducting the simulations. This research was partially supported by the ERC CoG 863818 (ForM-SMArt) grant and the Austrian Science Fund (FWF) 10.55776/COE12 grant.

\section*{Author Contributions}
A.K. contributed to the model formulation, developed the theoretical framework, and performed all simulations. 
K.K. supervised the project, conceived the study, contributed to the model formulation and theoretical analysis, and interpreted the results. 
K.C. provided theoretical analysis and contributed to the supervision and writing and editing of the manuscript. 
All authors discussed the results and contributed to the final manuscript.

\section*{Code and Data Availability}

The datasets generated during this study are available via the Ocean repository, as required by the journal. The simulation code and processed data necessary to reproduce the results are publicly available on Zenodo~\cite{zenodo_data}.

\clearpage
\appendix

    \begin{center}
        Supplementary Information for: \\
        \textbf{Predicting success of cooperators across arbitrary heterogeneous environmental landscapes} \\
        Amir Kargaran, Kamran Kaveh, Krishnendu Chatterjee
    \end{center}
    
    \vspace{1em}
    This PDF contains Supplementary Notes 1–6 and Supplementary Figures S1–S10. 
    
    \newpage

\setcounter{figure}{0}
\renewcommand{\thefigure}{S\arabic{figure}}

    
    \section*{Supplementary Note 1: Simplified Prisoner's Dilemma}\label{supp:note1}
    In our analysis, we use a simplified Prisoner's Dilemma (or pairwise cooperation game) with two strategies: cooperators (C) and defectors (D). A cooperator provides a benefit $b$ to its interaction partner while paying a cost $c$, whereas a defector provides no benefit and pays no cost. The resulting payoff matrix is
    \begin{equation}
        \label{eq:matrix}
        \begin{array}{c|cc}
             & C & D \\
            \hline
            C & b - c & -c \\
            D & b & 0 \\
        \end{array}
        \tag{S1}
    \end{equation}
    Fitness is determined from payoffs using a linear mapping. For illustration, consider a single pairwise interaction. If the focal individual plays against a cooperator or defector, its fitness is
    \begin{equation}
        \begin{aligned}
            C\!-\!C &\;\rightarrow\; F_{C} = 1 - w + w(b - c),\\
            C\!-\!D &\;\rightarrow\; F_{C} = 1 - w + w(-c),\\
            D\!-\!C &\;\rightarrow\; F_{D} = 1 - w + w(b),\\
            D\!-\!D &\;\rightarrow\; F_{D} = 1 - w.
        \end{aligned}
        \tag{S2}
    \end{equation}
    Here, $w$ denotes the selection intensity ($0 \le w \le 1$). Under weak selection ($w \to 0$), payoffs have only a small influence on fitness, whereas under strong selection ($w \to 1$), fitness is dominated by the game payoffs (Supplementary Equation~\ref{eq:matrix}). When $w = 0$, all individuals have fitness $1$ and the game has no effect on evolutionary dynamics.

	\newpage
	\section*{Supplementary Note 2: Exact fixation probability for the checkerboard configuration}\label{supp:note2}
    Here we extend the analysis of Ohtsuki et al.~\cite{Ohtsuki2006Royal} and derive the exact fixation probability of cooperators in a checkerboard environment under the Moran death-birth process~\cite{Moran_1958}. The heterogeneous payoff matrix for the simplified two-player cooperation game (Supplementary Note \hyperref[supp:note1]{1}) is shown in the Supplementary Table~\ref{tab:payoff_matrix} and includes all possible interactions when each player resides in a rich or poor site. For compactness, we define:
    \begin{equation}
        \begin{aligned}
            a_{\mathrm{r}} = (b_{\mathrm{ave}}+\sigma) - c,& \qquad a_{\mathrm{p}} = (b_{\mathrm{ave}}-\sigma) - c,&\\
            c_{\mathrm{r}} = b_{\mathrm{ave}}+\sigma,& \qquad c_{\mathrm{p}} = b_{\mathrm{ave}}-\sigma,&\\
            b=-c&,\qquad d=0.&
        \end{aligned}
        \tag{S9}
    \end{equation}
    These expressions give the payoff earned by the first player when interacting with the second player, whose strategy and site type are indicated by the matrix entries.
    \begin{table}[t]
        \centering
        \begin{tabular}{lcc|cc}
        \toprule
        \multicolumn{1}{c}{\textbf{First player}} & \multicolumn{4}{c}{\textbf{Second player}} \\
        \cmidrule(lr){2-5}
        & \multicolumn{2}{c}{\textbf{Rich site}} & \multicolumn{2}{c}{\textbf{Poor site}} \\
        \cmidrule(lr){2-3}\cmidrule(lr){4-5}
        & \textbf{Cooperate} & \textbf{Defect} & \textbf{Cooperate} & \textbf{Defect} \\
        \midrule
        \textsf{Rich site} \\
        \quad Cooperate & $a_{\mathrm{r}}$ & $b$ & $a_{\mathrm{p}}$ & $b$ \\
        \quad Defect     & $c_{\mathrm{r}}$ & $d$ & $c_{\mathrm{p}}$ & $d$ \\
        \midrule
        \textsf{Poor site} \\
        \quad Cooperate & $a_{\mathrm{r}}$ & $b$ & $a_{\mathrm{p}}$ & $b$ \\
        \quad Defect     & $c_{\mathrm{r}}$ & $d$ & $c_{\mathrm{p}}$ & $d$ \\
        \bottomrule
        \end{tabular}
        \caption{Payoff matrix for the simplified cooperation game with bimodal resource heterogeneity. Entries denote the payoff earned by the first player when interacting with the second player.}
        \label{tab:payoff_matrix}
    \end{table}
    We denote the number of cooperators by $i$ and the number of defectors by $N-i$. The transition probabilities
    \begin{equation}
        \lambda_i = \Pr(i \to i+1), \qquad \mu_i = \Pr(i \to i-1),
        \tag{S10}
    \end{equation}
    fully determine the stochastic dynamics. The exact fixation probability of cooperators~\cite{karlin1975first} is
    \begin{equation}\label{exact-fixation}
        \rho_C = \frac{1}{\,1 + \displaystyle\sum_{j=1}^{N-1} \prod_{i=1}^{j} \frac{\mu_i}{\lambda_i}\,}.
        \tag{S11}
    \end{equation}
    The ratio of fixation probabilities satisfies the following: 
    \begin{equation}
        \frac{\rho_C}{\rho_D}= \prod_{i=1}^{N-1}\frac{\lambda_i}{\mu_i}.
        \tag{S12}
    \end{equation}
    
    Thus, the key step is computing $\mu_i/\lambda_i$. To illustrate, panels (a) and (b) of the Supplementary Figure~\hyperref[fig:S1]{S1} show transitions for $i=1$ and $i=2$.
    \begin{figure}[p]
    	\centering
    	\includegraphics[width=\linewidth]{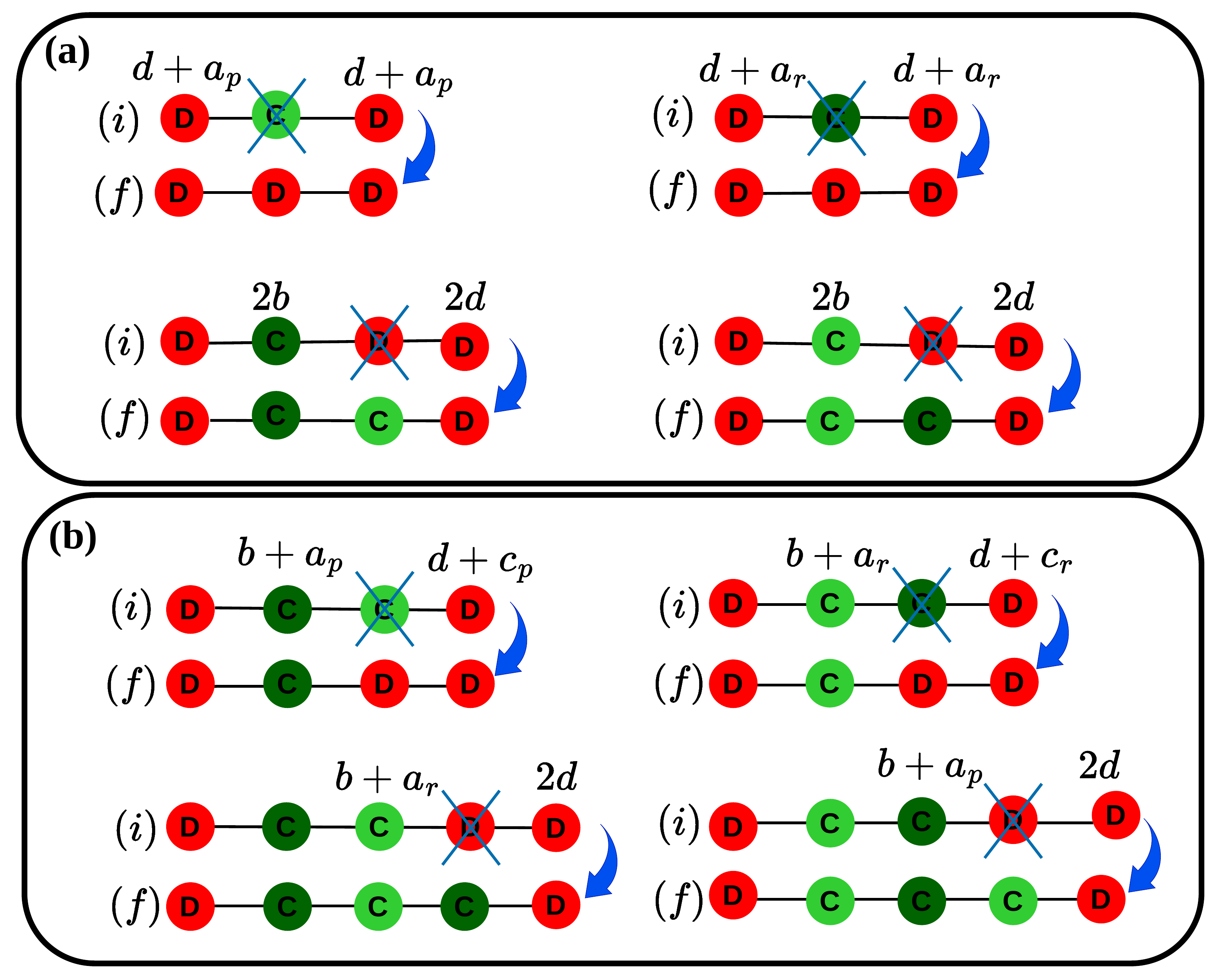}
    	\caption{
    		(a) Transitions for $i=1$: one cooperator going to zero or two.  
    		The payoffs of neighbors of the deceased individual are shown above each site.  
    		(b) Transitions for $i=2$: two cooperators going to one (top) or three (bottom).  
    		These cases determine $\mu_1$, $\lambda_1$, $\mu_2$, and $\lambda_2$ under the death--birth update rule.}
    	\label{fig:S1}
    \end{figure}
    To simplify expressions, we define:
    \begin{equation}
        \begin{aligned}
            \alpha_{\mathrm{p}}   &= 1-w + w(2 a_{\mathrm{p}}),   &\qquad \alpha_{\mathrm{r}}   &= 1-w + w(2 a_{\mathrm{r}}), \\[2mm]
            \gamma_{\mathrm{p}}   &= 1-w + w(2 c_{\mathrm{p}}),   &\qquad \gamma_{\mathrm{r}}   &= 1-w + w(2 c_{\mathrm{r}}), \\[2mm]
            \epsilon_{\mathrm{p}} &= 1-w + w(b + a_{\mathrm{p}}), &\qquad \epsilon_{\mathrm{r}} &= 1-w + w(b + a_{\mathrm{r}}), \\[2mm]
            \zeta_{\mathrm{p}}    &= 1-w + w(d + c_{\mathrm{p}}), &\qquad \zeta_{\mathrm{r}}    &= 1-w + w(d + c_{\mathrm{r}}), \\[3mm]
           & &\beta  = 1-w + w(2b),&\\[1mm]
           & &\delta = 1-w + w(2d).&
        \end{aligned}
        \tag{S13}
    \end{equation}
    \noindent\textbf{Case \(i = 1\).} A population with one cooperator reaches $0$ cooperators with probability:
    \begin{equation}
        \mu_1 = \frac{1}{N}.
        \tag{S14}
    \end{equation}    
    To increase to two cooperators, the new cooperator can appear on either side of the lone cooperator.  
    The reproduction probability is \(F_C/(F_C+F_D)\), and averaging over rich/poor sites gives:
    \begin{equation}
        \lambda_1 = \frac{2}{N} \frac{\beta}{\beta + \delta}.
        \tag{S15}
    \end{equation}
    Thus:
    \begin{equation}
        \frac{\mu_1}{\lambda_1} = \frac{\beta+\delta}{2\beta}.
        \tag{S16}
    \end{equation}
    \medskip
    \noindent\textbf{Case \(i = 2\).} From the Supplementary Figure~\hyperref[fig:S1]{S1} (b), we compute:
    \begin{equation}
        \mu_2= \frac{1}{2N}\left(\frac{\zeta_{\mathrm{p}}}{\zeta_{\mathrm{p}}+\epsilon_{\mathrm{p}}}+\frac{\zeta_{\mathrm{r}}}{\zeta_{\mathrm{r}}+\epsilon_{\mathrm{r}}}\right),
        \lambda_2 = \frac{1}{2N}\left(\frac{\epsilon_{\mathrm{r}}}{\epsilon_{\mathrm{r}}+\delta}+\frac{\epsilon_{\mathrm{p}}}{\epsilon_{\mathrm{p}}+\delta}\right).
        \tag{S17}
    \end{equation} 
    Thus:
    \begin{equation}
        \frac{\mu_2}{\lambda_2}=\frac{\frac{\zeta_{\mathrm{p}}}{\zeta_{\mathrm{p}}+\epsilon_{\mathrm{p}}}+\frac{\zeta_{\mathrm{r}}}{\zeta_{\mathrm{r}}+\epsilon_{\mathrm{r}}}}{\frac{\epsilon_{\mathrm{r}}}{\epsilon_{\mathrm{r}}+\delta}
        +\frac{\epsilon_{\mathrm{p}}}{\epsilon_{\mathrm{p}}+\delta}}.
        \tag{S18}
    \end{equation}
    \medskip
    \noindent\textbf{General case \(3 \le i \le N-1\).} Carrying out analogous calculations yields:
    \begin{equation}\label{TC}
        \frac{\mu_i}{\lambda_i} =
            \begin{cases}
                \dfrac{\beta+\delta}{2\beta}, & i = 1,\\[3mm]
                \dfrac{
                \frac{\zeta_{\mathrm{p}}}{\zeta_{\mathrm{p}}+\epsilon_{\mathrm{p}}}
                +
                \frac{\zeta_{\mathrm{r}}}{\zeta_{\mathrm{r}}+\epsilon_{\mathrm{r}}}
                }{
                \frac{\epsilon_{\mathrm{r}}}{\epsilon_{\mathrm{r}}+\delta}
                +
                \frac{\epsilon_{\mathrm{p}}}{\epsilon_{\mathrm{p}}+\delta}
                }, 
                & i = 2,\\[4mm]
                \dfrac{
                \frac{\zeta_{\mathrm{r}}}{\zeta_{\mathrm{r}}+\alpha_{\mathrm{r}}}
                +
                \frac{\zeta_{\mathrm{p}}}{\zeta_{\mathrm{p}}+\alpha_{\mathrm{p}}}
                }{
                \frac{\epsilon_{\mathrm{r}}}{\epsilon_{\mathrm{r}}+\delta}
                +
                \frac{\epsilon_{\mathrm{p}}}{\epsilon_{\mathrm{p}}+\delta}
                },
                & 3 \le i \le N-3, \\[4mm]
                \dfrac{
                \frac{\zeta_{\mathrm{r}}}{\zeta_{\mathrm{r}}+\alpha_{\mathrm{r}}}
                +
                \frac{\zeta_{\mathrm{p}}}{\zeta_{\mathrm{p}}+\alpha_{\mathrm{p}}}
                }{
                \frac{\epsilon_{\mathrm{r}}}{\zeta_{\mathrm{r}}+\epsilon_{\mathrm{r}}}
                +
                \frac{\epsilon_{\mathrm{p}}}{\zeta_{\mathrm{p}}+\epsilon_{\mathrm{p}}}
                },
                & i = N-2,\\[4mm]
                \dfrac{\gamma_{\mathrm{r}}}{\gamma_{\mathrm{r}}+\alpha_{\mathrm{r}}}
                +
                \dfrac{\gamma_{\mathrm{p}}}{\gamma_{\mathrm{p}}+\alpha_{\mathrm{p}}},
                & i = N-1.
            \end{cases}
            \tag{S19}
    \end{equation}
    When $a_{\mathrm{p}} = a_{\mathrm{r}}$ and $c_{\mathrm{p}} = c_{\mathrm{r}}$, this reproduces the uniform-environment results of Ohtsuki et al.~\cite{Ohtsuki2006Royal}.
    Supplementary Figure~\hyperref[fig:S2]{S2} compares the analytical result from Supplementary Equation~\ref{exact-fixation} and Supplementary Equation~\ref{TC} with simulations.
	\begin{figure}[p]
		\centering
		\includegraphics[width=\linewidth]{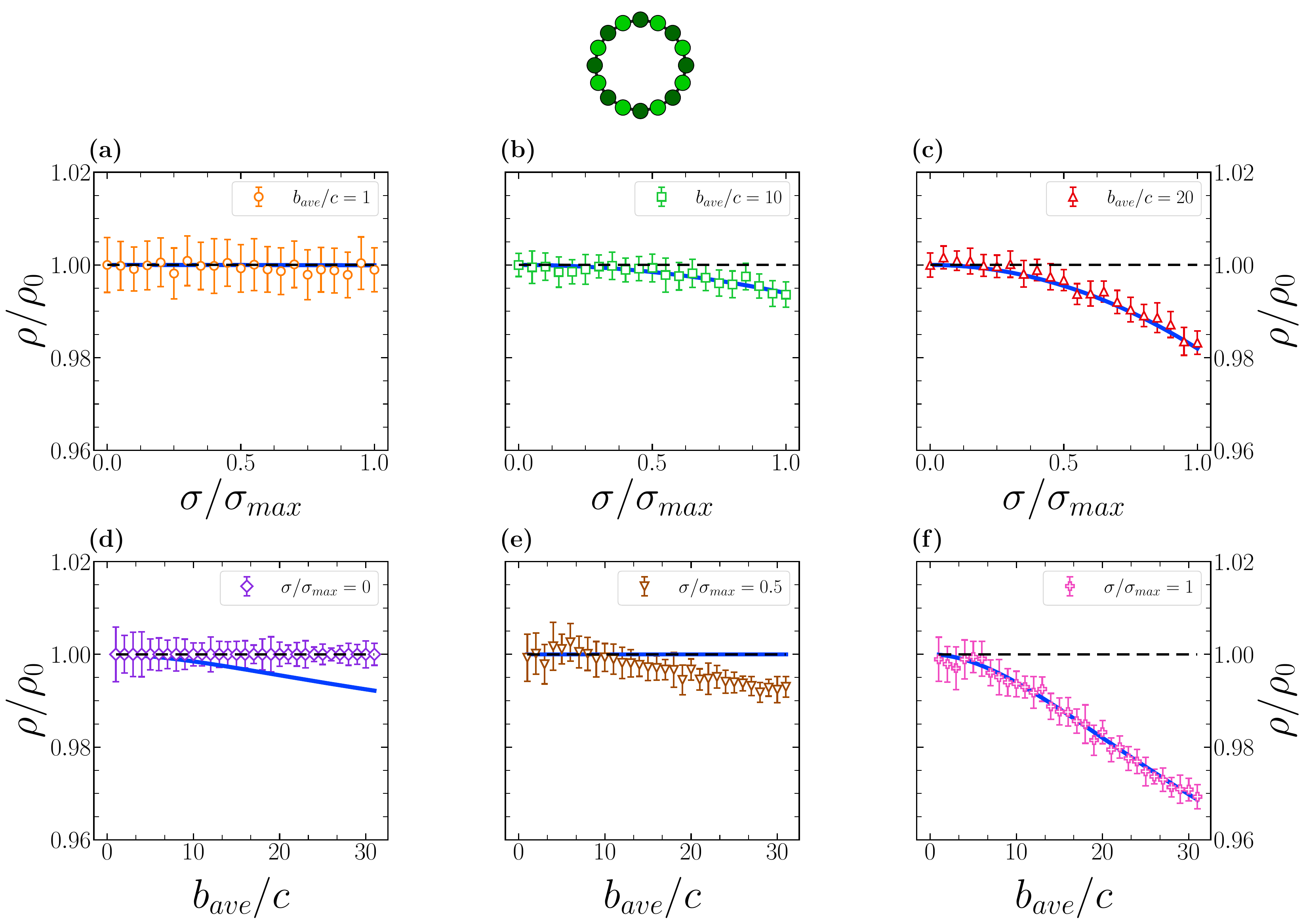}
		\caption{Comparison of analytical fixation probability (Supplementary Equation~\ref{exact-fixation}, blue solid curve) with simulations for the  checkerboard configuration. Panels (a--c): varying $b_{\mathrm{ave}}/c$ at fixed $\sigma/\sigma_{\max}$. Panels (d--f): varying $\sigma/\sigma_{\max}$ at fixed $b_{\mathrm{ave}}/c$. Agreement is excellent at fixed benefit-to-cost ratio; at fixed heterogeneity, accuracy is highest when $\sigma/\sigma_{\max} = 1$. Parameters: $c = 0.125$, $N = 100$, $w = 0.01$. Insets show 16-node networks for clarity.}
		\label{fig:S2}
	\end{figure}
	\clearpage
\begin{figure}[p]
	\centering
	\includegraphics[width=\linewidth]{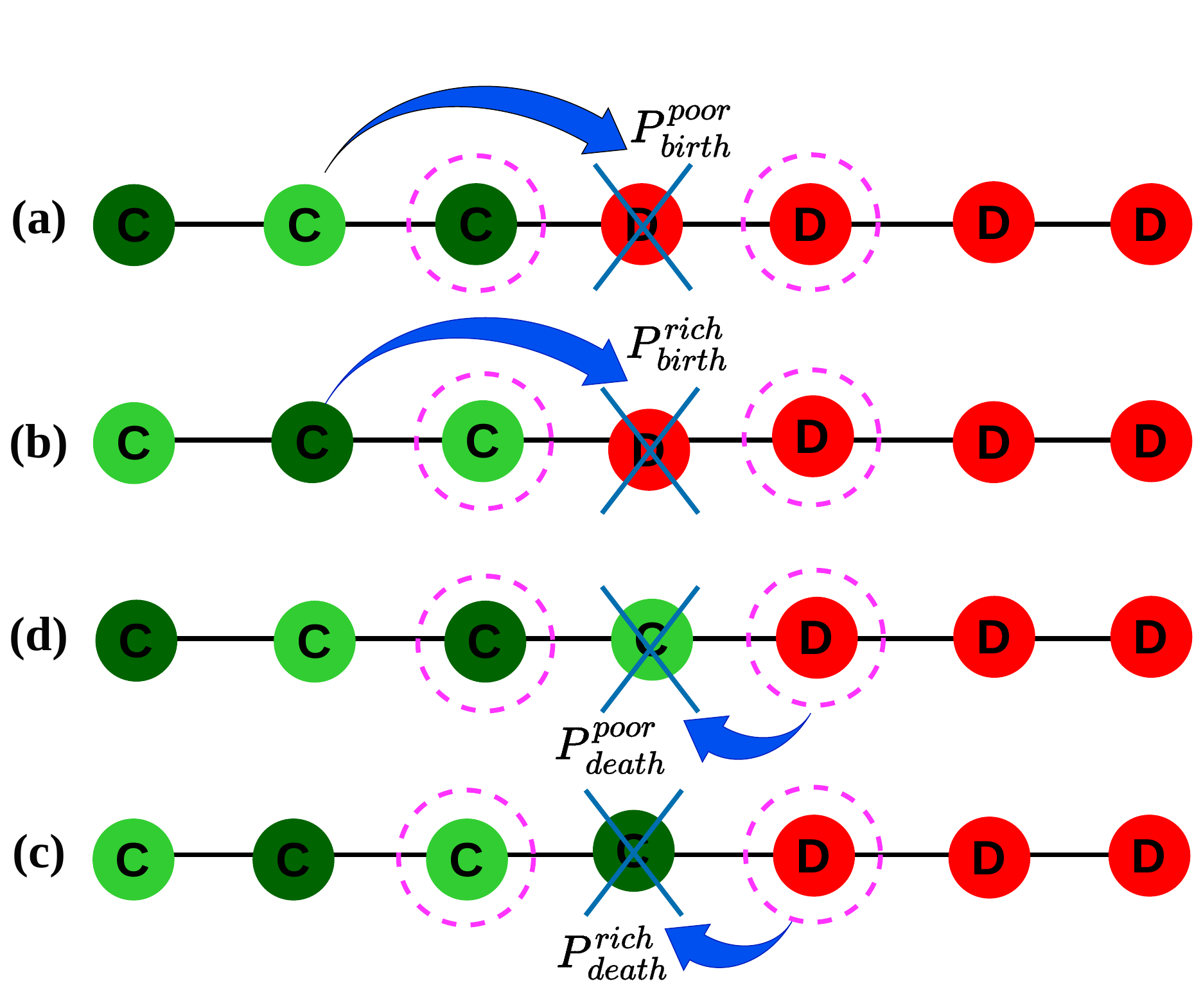}
	\caption{Update events in a checkerboard configuration on a cycle. (a, b) A defector dies and is replaced by a poor (light green) or rich (dark green) cooperator with probabilities $P_{\mathrm{birth}}^{\mathrm{poor}}$ and $P_{\mathrm{birth}}^{\mathrm{rich}}$. (c, d) A poor or rich cooperator dies and is replaced by a defector with probabilities $P_{\mathrm{death}}^{\mathrm{poor}}$ and $P_{\mathrm{death}}^{\mathrm{rich}}$. Reproduction probabilities depend on the fitness of the two neighbors of the deceased individual.}
	\label{fig:S3}
\end{figure}
    \section*{Supplementary Note 3: Derivation of the \texorpdfstring{\textit{b/c}}{b/c}-rule for the checkerboard configuration}\label{supp:note3}
 
    Here we derive the \textit{b/c}-rule for the checkerboard configuration on a cycle graph using the death--birth Moran process \cite{Moran_1958}. The Supplementary Figure~\hyperref[fig:S3]{S3} illustrates all possible update events that increase or decrease the number of cooperators. If a defector dies, its two neighbors compete to fill the vacancy, and a poor or rich cooperator may be born.  
    We denote these probabilities by $P_{\mathrm{birth}}^{\mathrm{poor}}$ and $P_{\mathrm{birth}}^{\mathrm{rich}}$. Conversely, if a cooperator dies, a poor or a rich one, a defector may be born with probabilities $P_{\mathrm{death}}^{\mathrm{poor}}$ and $P_{\mathrm{death}}^{\mathrm{rich}}$. All of these are derived from the standard fitness-weighted reproduction rule. As an example, consider panel (a) of the Supplementary Figure~\hyperref[fig:S3]{S3}. A defector dies, leaving two neighbors to compete: a poor cooperator and a defector. Their fitness are 
    \begin{equation}
        F_C = 1 - w + w(b_{\mathrm{poor}} - 2c), \qquad F_D = 1 - w,
        \tag{S3}
    \end{equation}
    leading to the birth probability;
    \begin{equation}
        P_{\mathrm{birth}}^{\mathrm{poor}}= \frac{F_C}{F_C + F_D} = \frac{1-w + w(b_{\mathrm{poor}} - 2c)} {2(1-w) + w(b_{\mathrm{poor}} - 2c)}.
        \tag{S4}
    \end{equation}
    Proceeding analogously for all cases yields the following:
    \begin{equation}
        \begin{aligned}
            P_{\mathrm{birth}}^{\mathrm{rich}}
            &= \frac{1-w + w(b_{\mathrm{rich}} - 2c)}
                     {2(1-w) + w(b_{\mathrm{rich}} - 2c)}, \\
            P_{\mathrm{death}}^{\mathrm{poor}}
            &= \frac{1-w + w b_{\mathrm{poor}}}
                     {2(1-w) + w(3b_{\mathrm{poor}} - 2c)}, \\
            P_{\mathrm{death}}^{\mathrm{rich}}
            &= \frac{1-w + w b_{\mathrm{rich}}}
                     {2(1-w) + w(3b_{\mathrm{rich}} - 2c)}.
        \end{aligned}
        \tag{S5}
    \end{equation}
    The probability that the number of cooperators increases from $n$ to $n+1$ is the probability that a defector dies ($1/N$ on the cycle) multiplied by the probability that either type of cooperator is born:
    \begin{equation}
        \begin{aligned}
            \mathcal{P}^+(n\to n+1)
            &= \frac{1}{N}
            \bigg[
            \frac{1-w + w(b_{\mathrm{poor}} - 2c)}
                 {2(1-w) + w(b_{\mathrm{poor}} - 2c)}
            +
            \frac{1-w + w(b_{\mathrm{rich}} - 2c)}
                 {2(1-w) + w(b_{\mathrm{rich}} - 2c)}
            \bigg].
        \end{aligned}
        \tag{S6}
    \end{equation}
    Similarly, the probability of a decrease ($n \to n-1$) is:
    \begin{equation}
        \begin{aligned}
            \mathcal{P}^-(n\to n-1)&= \frac{1}{N}\bigg[\frac{1-w + w b_{\mathrm{poor}}}{2(1-w) + w(3b_{\mathrm{poor}} - 2c)}+\frac{1-w + w b_{\mathrm{rich}}}{2(1-w) + w(3b_{\mathrm{rich}} - 2c)}\bigg].
        \end{aligned}
        \tag{S7}
    \end{equation}
    A cooperator is favored to fix if $\mathcal{P}^+ > \mathcal{P}^-$. Under weak selection ($w \to 0$), expanding to the first order yields
    \begin{equation}\label{eq:checker_bc_rule}
        \mathcal{P}^+ > \mathcal{P}^- \quad \Longrightarrow \quad(b_{\mathrm{poor}} + b_{\mathrm{rich}} - 4c)N w > 0\quad \Longrightarrow \quad\frac{b_{\mathrm{ave}}}{c} > 2.
        \tag{S8}
    \end{equation}

    Thus, the \textit{b/c}-rule for the checkerboard configuration is identical to the uniform case at $\mathcal{O}(w)$, demonstrating that heterogeneity affects only higher-order terms in the weak-selection expansion.

 	\section*{Supplementary Note 4: Approximating fixation probability for segregated configurations}\label{supp:note4}
\begin{figure}[p]
		\centering
		\includegraphics[width=\linewidth]{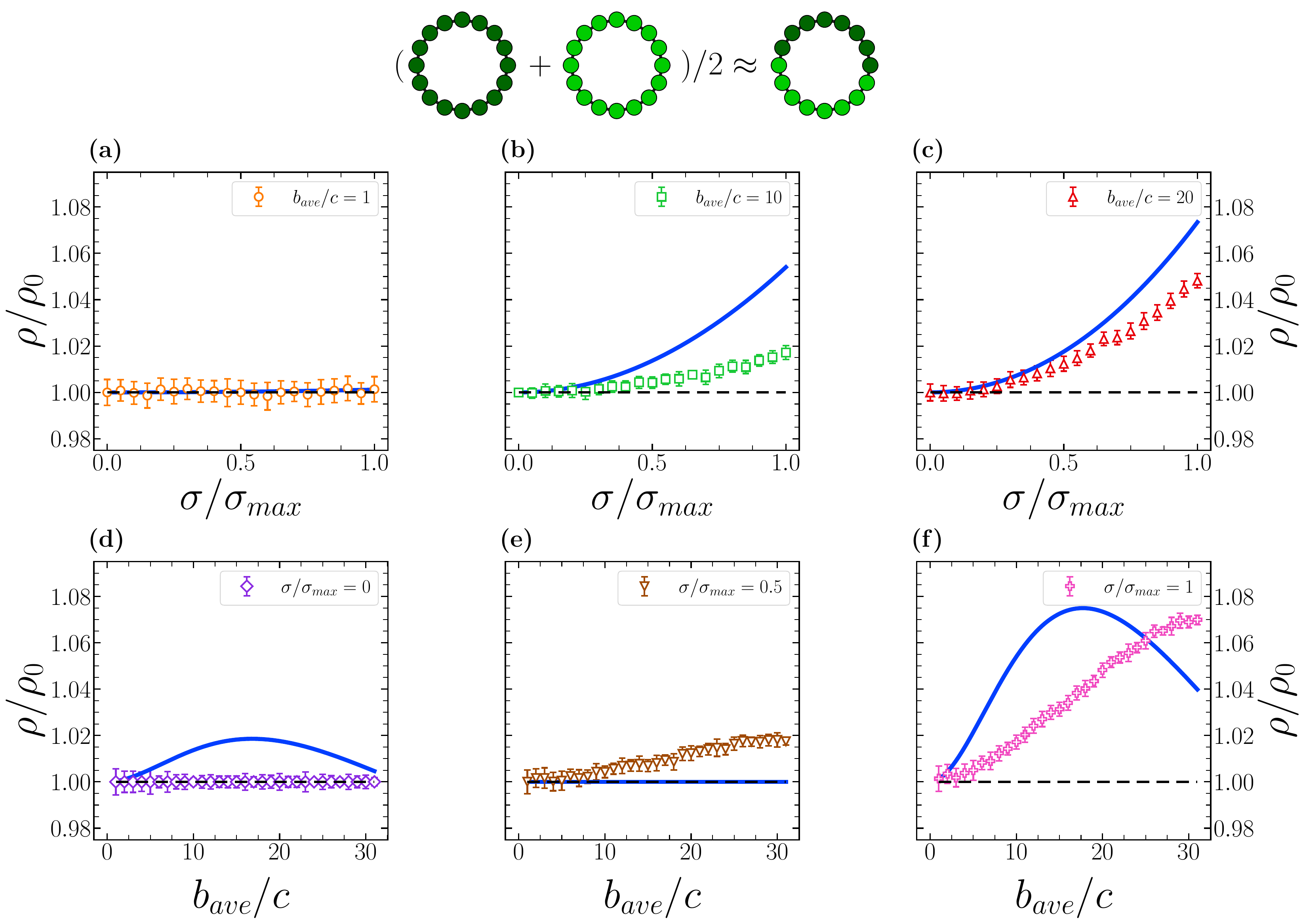}
		\caption{Comparison of analytical fixation probability for the segregated configuration (Supplementary Equation~\ref{segfix}, blue solid curve) with simulations. Panels (a--c): varying $b_{\mathrm{ave}}/c$. Panels (d-f): varying $\sigma/\sigma_{\max}$. Agreement is strong for fixed $b_{\mathrm{ave}}/c$, but weaker when sweeping heterogeneity. Parameters: $c = 0.125$, $N = 100$, $w = 0.01$. Insets show 16 nodes for clarity.}
		\label{fig:S4}
	\end{figure}

    Here, we approximate the fixation probability for the segregated configuration. We rely on the exact result of Ohtsuki et al.~\cite{Ohtsuki2006Royal} for the death-birth Moran process on a cycle with two neighbors (Supplementary Equation~\ref{TC} and Supplementary Equation~\ref{exact-fixation} when $a_{\mathrm{p}} = a_{\mathrm{r}}$ and $c_{\mathrm{p}} = c_{\mathrm{r}}$). We treat this exact fixation probability as a function:
    \begin{equation}
        \rho_{\mathrm{C}} = f(a,b,c,d,w,N). 
        \tag{S20}
    \end{equation}
    Assuming two separate cycles of size \(N\), one with all rich cooperators and one with all poor cooperators, we approximate the fixation probability in the segregated configuration by averaging the corresponding fixation probabilities. Thus,
    \begin{equation}\label{segfix}
        \rho_{\mathrm{C}}^{\mathrm{S}}
        \approx
        \frac{
        f(a_{\mathrm{p}}, b, c_{\mathrm{p}}, d, w, N)
        +
        f(a_{\mathrm{r}}, b, c_{\mathrm{r}}, d, w, N)
        }{2},
        \tag{S21}
    \end{equation}
    where
     \begin{equation}
         \begin{aligned}
            a_{\mathrm{r}} = (b_{\mathrm{ave}}+\sigma)-c,&\qquad
            a_{\mathrm{p}} = (b_{\mathrm{ave}}-\sigma)-c,&\\
            c_{\mathrm{r}} = b_{\mathrm{ave}}+\sigma,&\qquad
            c_{\mathrm{p}} = b_{\mathrm{ave}}-\sigma,&
         \end{aligned}
         \tag{S22} 
    \end{equation}
    and we use \(b = -c\) and \(d = 0\). We validated this approximation through simulations (Supplementary Figure~\hyperref[fig:S4]{S4}) and successfully capture the qualitative trend of fixation probability in the segregated configuration. Under weak selection (\(w \ll 1\)), we expand the Supplementary Equation~\eqref{segfix} to obtain the corresponding \textit{b/c} rule:
    \begin{equation}\label{segsimplerule}
        \rho_{\mathrm{C}}^{\mathrm{S}} > \frac{1}{N}
        \;\Longrightarrow\;
        \frac{
        w \left[ b_{\mathrm{ave}} (N-4) - 2c (N-2) \right]
        }{2N}
        > 0
        \;\Longrightarrow\;
        \frac{b_{\mathrm{ave}}}{c}
        >
        \frac{2 (N-2)}{N-4}.
        \tag{S23}
    \end{equation}
    For large population sizes (\(N \gg 1\)), this condition approaches
    \begin{equation}
        \frac{b_{\mathrm{ave}}}{c} > 2.
        \tag{S24}
    \end{equation}
    The effect of heterogeneity appears only in higher-order terms of the weak-selection expansion; it does not modify the first-order \(\mathcal{O}(w)\) rule expressed above.
    \newpage
    \begin{figure}
        \centering
        \includegraphics[width=1\linewidth]{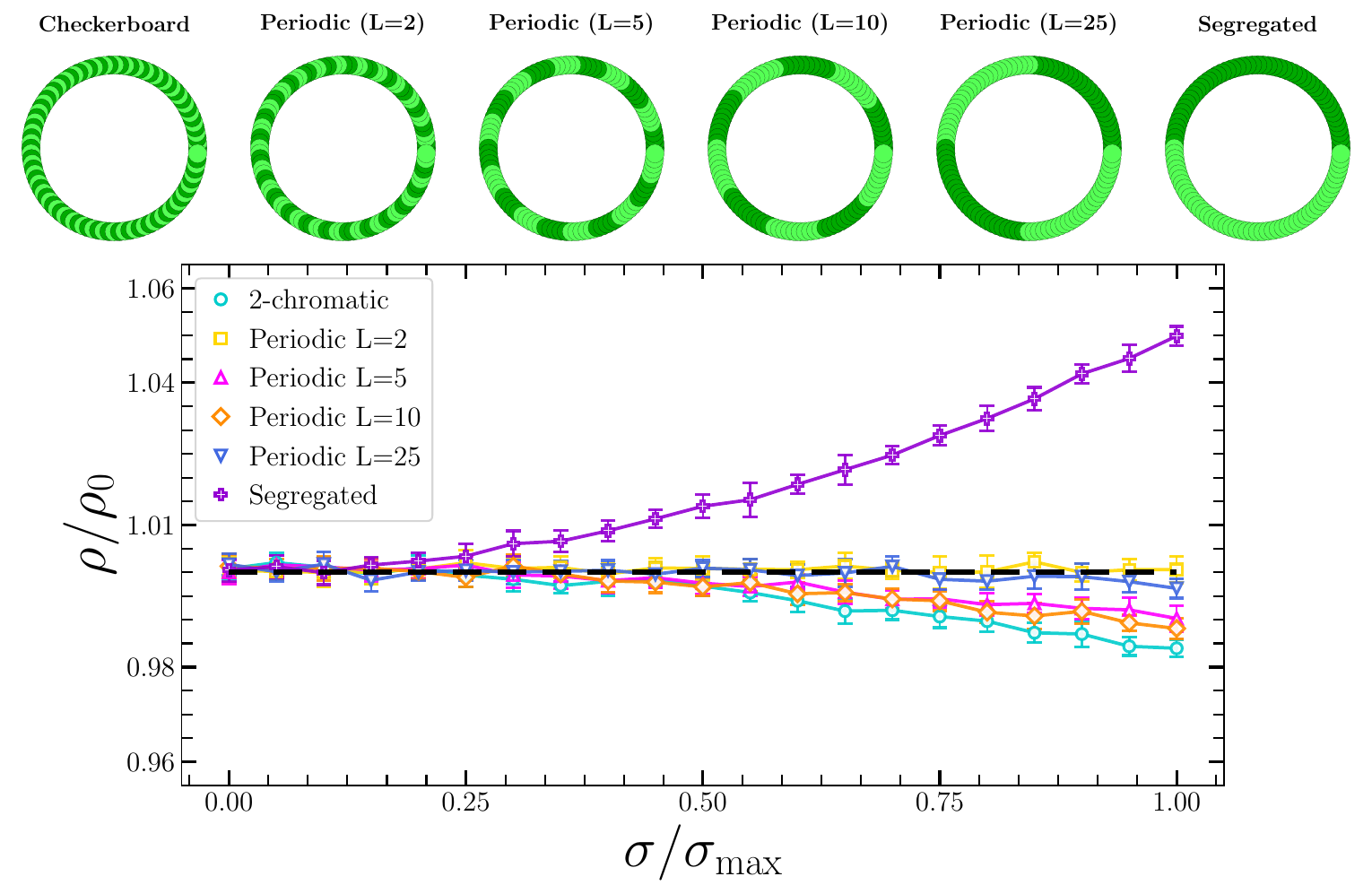}
        \caption{Normalized fixation probability ($\rho/\rho_0$) versus normalized heterogeneity parameter ($\sigma/\sigma_{\text{max}}$) for periodic configurations on the cycle graph. Here, “period” refers to the interval over which an equal number of rich (dark green) and poor (light green) cooperators repeat uniformly around the cycle. Parameters: $b_{\text{ave}}/c=20$, $N=100$, $w=0.01$, with an equal number of poor and rich sites in all configurations.}
        \label{fig:S5}
    \end{figure}
    
    \begin{figure}[p]
        \centering
        \includegraphics[width=1\linewidth]{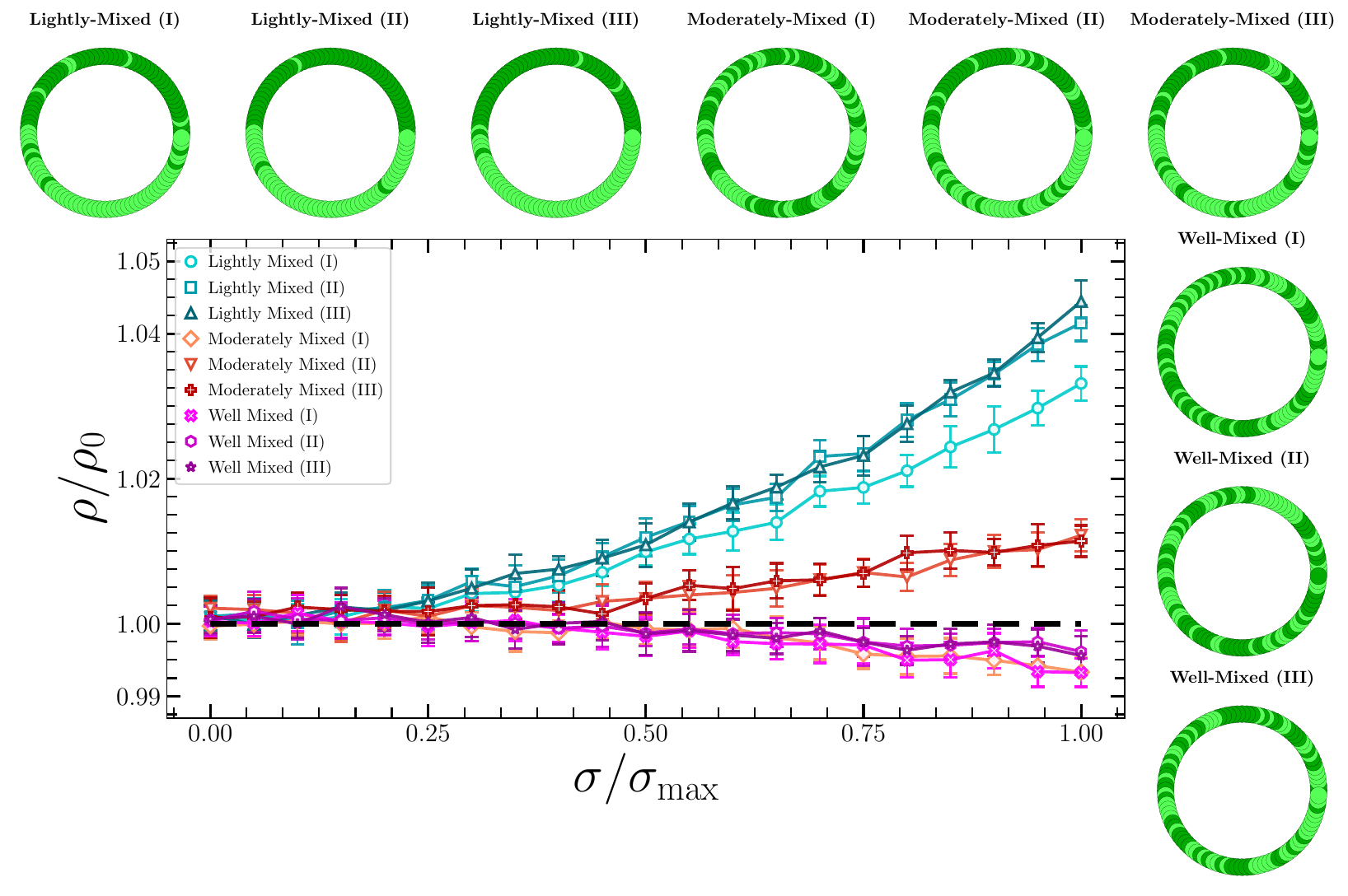}
        \caption{Normalized fixation probability ($\rho/\rho_0$) versus normalized heterogeneity ($\sigma/\sigma_{\text{max}}$) for three classes of random configurations on the cycle: Lightly-Mixed, Moderately-Mixed, and Well-Mixed. Swapping numbers are drawn randomly from the ranges $1$–$10$, $10$–$10^2$, and $10^2$–$10^3$, respectively. Lightly-Mixed configurations resemble segregated patterns and therefore show increasing trends, whereas Moderately- and Well-Mixed configurations behave more like checkerboard patterns and show decreasing trends. Parameters: $b_{\mathrm{ave}}/c=20$, $N=100$, $w=0.01$, with an equal number of poor and rich sites in all configurations.}
        \label{fig:S6}
    \end{figure}

    \section*{Supplementary Note 5: Comparative analysis across patterned and random configurations}\label{supp:note5}
\begin{figure}[p]
        \centering
        \includegraphics[width=1\linewidth]{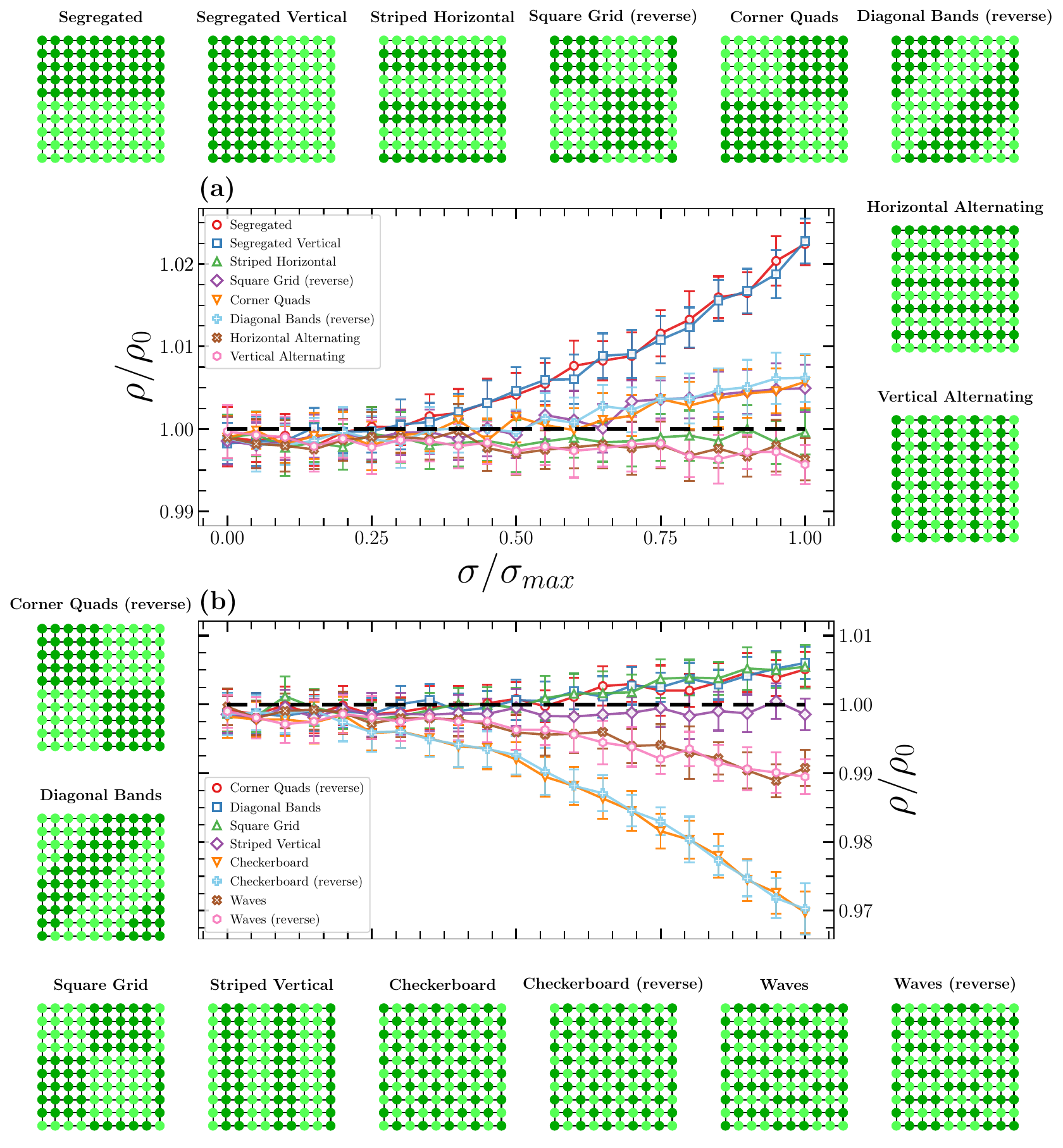}
        \caption{Normalized fixation probability ($\rho/\rho_0$) versus normalized heterogeneity ($\sigma/\sigma_{\text{max}}$) for a range of periodic and pattern-like configurations on a square lattice, spanning the full spectrum from segregated to checkerboard-like patterns. Configurations resembling the segregated pattern exhibit increasing trends, whereas those similar to checkerboard patterns show decreasing trends with heterogeneity. Parameters: $b_{\mathrm{ave}}/c=20$, $N=100$, $w=0.01$, with equal numbers of poor and rich sites.}
        \label{fig:S7}
    \end{figure}

\begin{figure}[p]
        \centering
        \includegraphics[width=1\linewidth]{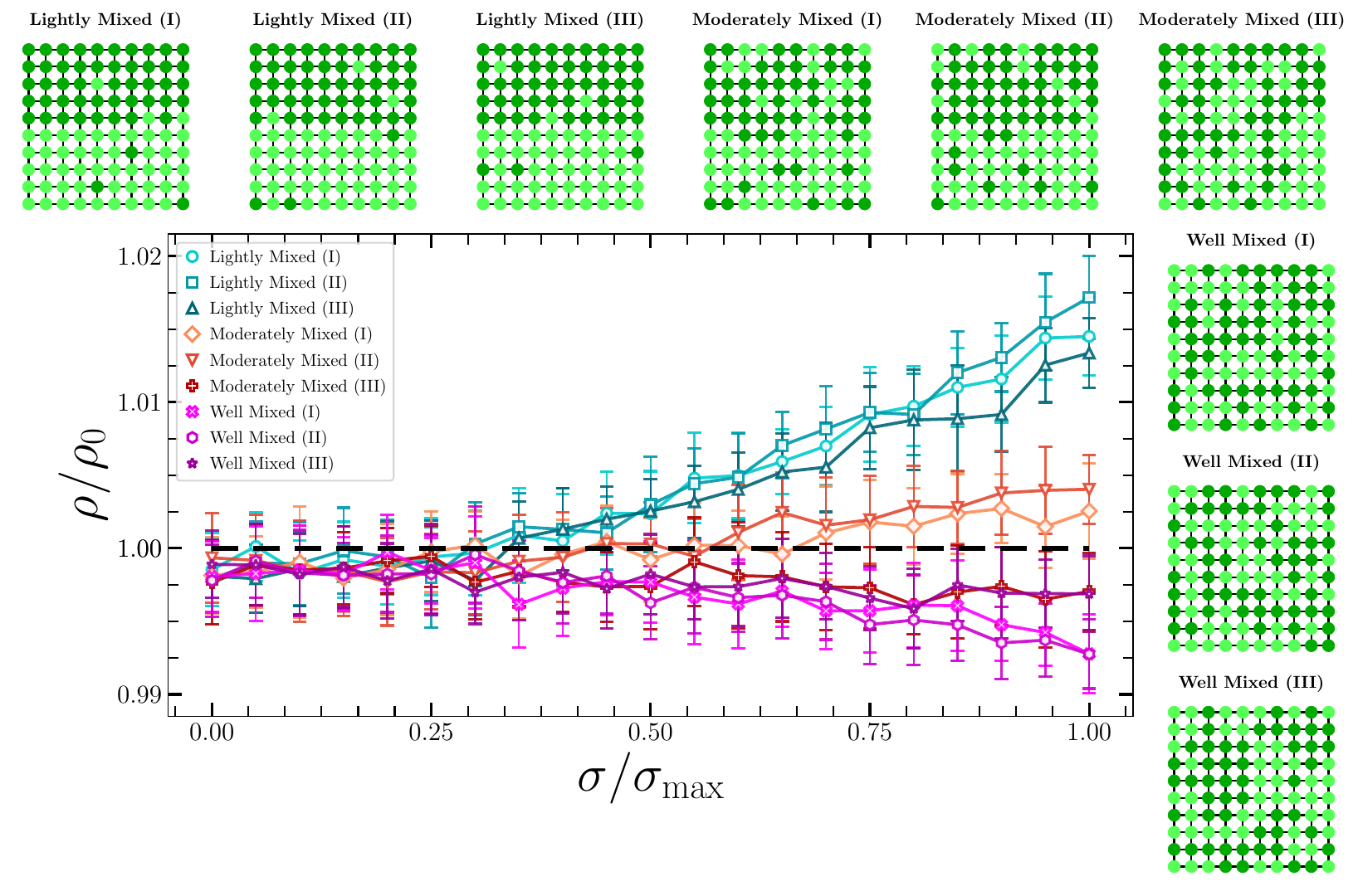}
        \caption{Normalized fixation probability ($\rho/\rho_0$) versus normalized heterogeneity ($\sigma/\sigma_{\text{max}}$) for random configurations on a square lattice. Configurations are generated by randomly swapping positions in a segregated pattern, with swapping numbers chosen from the ranges $1$–$10$ (Lightly-Mixed), $10$–$10^2$ (Moderately-Mixed), and $10^2$–$10^3$ (Well-Mixed). Lightly-Mixed configurations behave similarly to segregated patterns, while Moderately- and Well-Mixed configurations resemble checkerboard patterns. Parameters: $b_{\mathrm{ave}}/c=20$, $N=100$, $w=0.01$, with an equal number of poor and rich sites in all cases.}
        \label{fig:S8}
    \end{figure}

    In this section, we examine how the normalized fixation probability ($\rho/\rho_0$) varies with the heterogeneity parameter ($\sigma$) across 40 different configurations. These configurations include random, periodic, and pattern-like arrangements on both circular networks and square lattices. Supplementary figures~\hyperref[fig:S5]{S5} and Supplementary Figures~\hyperref[fig:S6]{S6} show the behavior of $\rho/\rho_0$ for periodic and random configurations on the 1D cycle, respectively. Supplementary Figure~\hyperref[fig:S7]{S7} shows the corresponding behavior for periodic or pattern-like configurations on 2D lattices. Finally, a Supplementary Figure~\hyperref[fig:S8]{S8} presents results for random configurations on square lattices.
    
    Across all configurations, we observe a consistent pattern: Configurations that closely resemble the segregated configuration exhibit an increasing normalized fixation probability as heterogeneity increases, whereas configurations similar to the checkerboard configuration show a decreasing trend. Random configurations fall anywhere along this spectrum, depending on the degree of mixing between rich and poor sites.
 
	\newpage
    \section*{Supplementary Note 6: Temporal dynamics of the level of cooperation}\label{supp:note6}
    In this section, we examine the temporal evolution of cooperation in checkerboard and segregated configurations for cycle graphs (panels (a) and (b)) and square lattices (panels (c) and (d)) in the Supplementary Figure~\hyperref[fig:S9]{S9}, which includes time-series panels and summary comparisons between system sizes. Each simulation begins from an initial state containing $10\%$ cooperators and evolves for a finite duration ($t \le 10^9$). The level of cooperation is computed by averaging over 100 realizations conditioned on non-extinction of cooperators up to time $t$. For cycle graphs of sizes $N=10^3$, $2\times10^3$, $5\times10^3$, and $10^4$, the number of realizations that reached full cooperation was $91$, $100$, $100$, and $100$, respectively, for the segregated configuration, and $89$, $99$, $100$ and $100$ for the checkerboard configuration.
    
    At the smallest size ($N=10^3$), both configurations achieve full cooperation within the simulated time. For cycle graphs of size $N=2\times 10^3$, the checkerboard configuration typically reaches the all-cooperator state within the time window ($t<10^9$), whereas the segregated configuration often fails to do so. For larger cycles ($N=5\times10^3$ and $10^4$), the segregated configuration rapidly settles into a metastable, partially cooperative steady state with an effectively zero growth rate, while the checkerboard configuration still reaches fixation within a finite time.
    
    A similar pattern appears in square lattices (panels (c) and (d)), although the timescales differ. For the lattice sizes $20\times 20$, $50\times 50$, $100\times 100$ and $150\times 150$, the numbers of realizations that reached maximal cooperation were $55$, $100$, $100$ and $100$ for the segregated configuration, and $56$, $99$, $100$, and $100$ for the checkerboard configuration. In segregated lattices, the dynamics rapidly freezes to a partially cooperative state with negligible growth, rendering full cooperation effectively unreachable. In contrast, the checkerboard configuration reaches full cooperation much more rapidly, typically by $t < 4\times 10^7$, and substantially faster than in the corresponding 1D systems.

 \begin{figure}[p]
    	\centering
    	\includegraphics[width=1\linewidth]{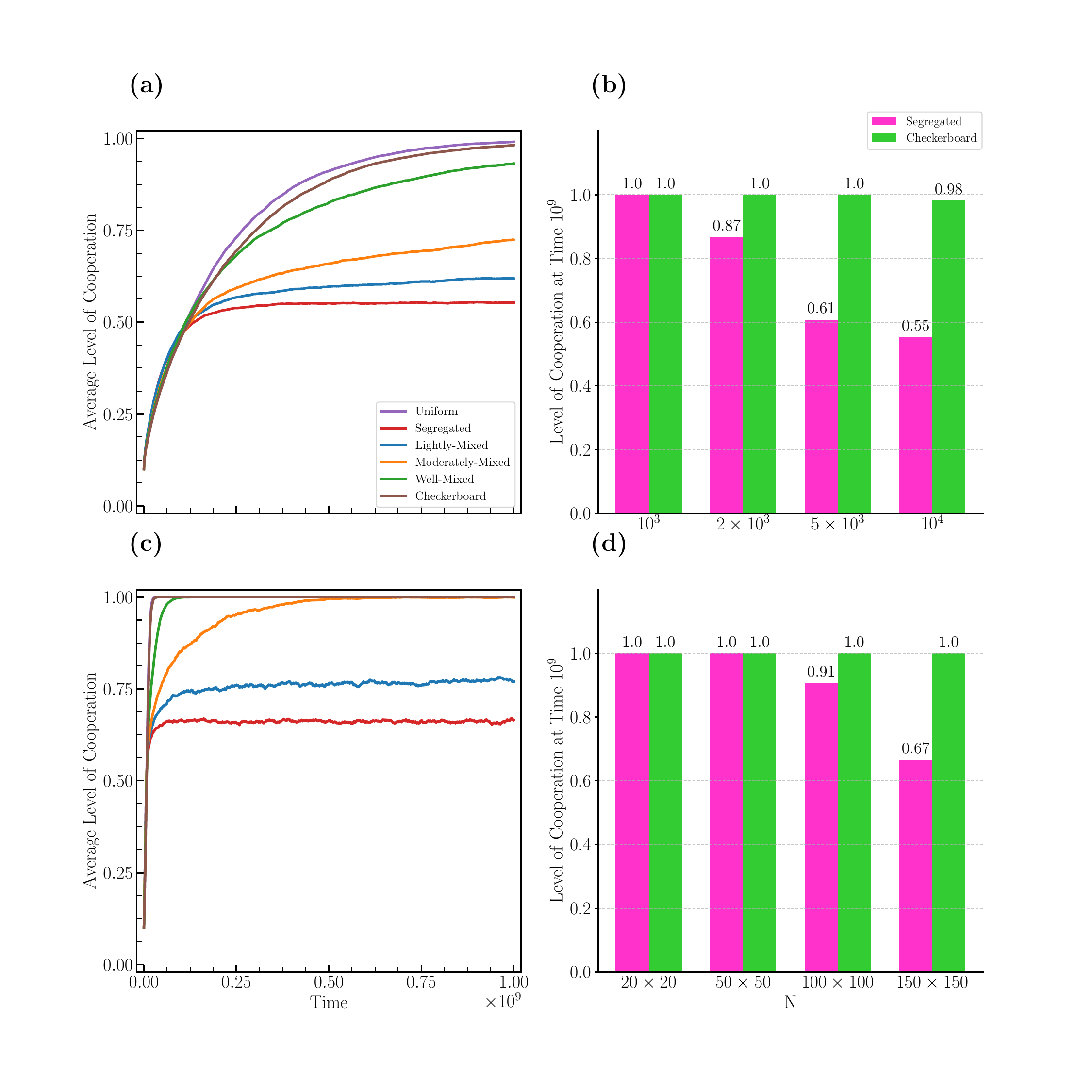}
    	\caption{Panel (a) shows results on a cycle graph with $N = 10^4$ nodes and panel (c) on a square lattice $150 \times 150$, both initialized with $10\%$ cooperators. The mean benefit is $b_{\mathrm{ave}} = 2.5$ and the normalized heterogeneity is $\sigma/\sigma_{\text{max}} = 1$. Random configurations are generated by applying $10^3$, $2 \times 10^3$, and $5 \times 10^3$ swaps to the segregated state. Panels (b) and (d) show bar charts comparing the final cooperation levels ($t = 10^9$) between the checkerboard and segregated configurations for multiple system sizes.}
    	\label{fig:S9}
    \end{figure}

    \begin{figure}[p]
    	\centering
    	\includegraphics[width=0.9\linewidth]{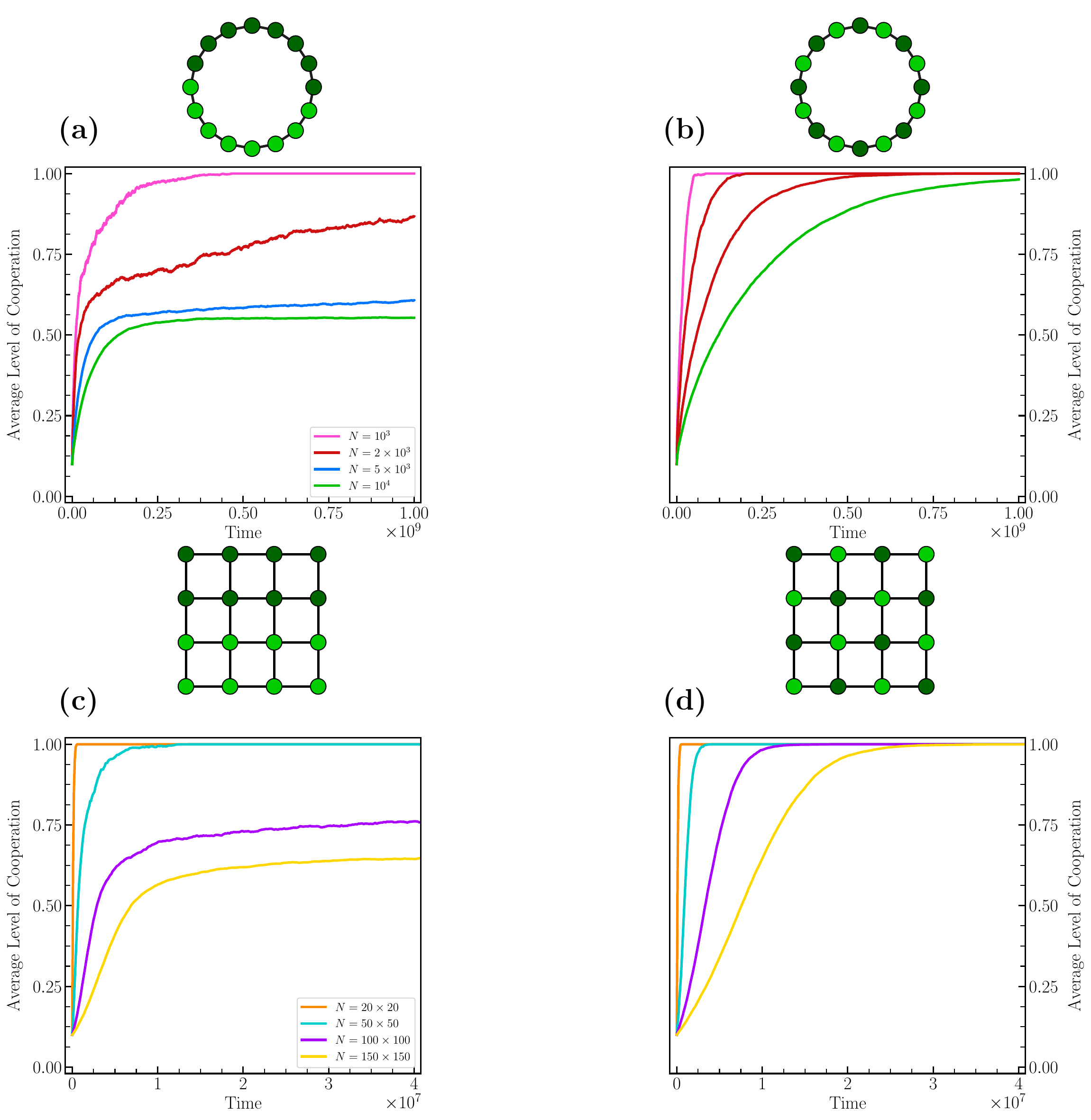}
    	\caption{Time evolution of the cooperation level for cycle graphs (top row) and square lattices (bottom row), comparing segregated (panels a and c) and checkerboard (panels b and d) configurations across multiple system sizes. The mean benefit is $b_{\mathrm{ave}} = 2.5$, and the heterogeneity parameter is $\sigma = 2.5$, corresponding to the strongest heterogeneity considered. For large cycle sizes ($N=5\times10^3$ and $10^4$), the segregated configuration quickly becomes trapped in a metastable partially cooperative state, while the checkerboard configuration eventually reaches full cooperation (by $t \approx 10^9$). In square lattices, segregated configurations freeze into a metastable state, whereas the checkerboard configuration achieves full cooperation much more rapidly ($t < 4\times10^7$).}
    	\label{fig:S10}
    \end{figure}

\end{document}